# Electronics and optoelectronics of quasi-one dimensional layered transition metal trichalcogenides

Joshua O. Island[1,†], Aday J. Molina-Mendoza[2,‡], Mariam Barawi[3,*], Robert Biele[4], Eduardo Flores[3], José M. Clamagirand[3], José R. Ares[3], Carlos Sánchez[3,5], Herre S.J. van der Zant[1], Roberto D'Agosta[4,6], Isabel J. Ferrer[3,5], Andres Castellanos-Gomez[7]

[1] Kavli Institute of Nanoscience, Delft University of Technology, Lorentzweg 1, 2628 CJ Delft, The Netherlands.

[2] Dpto. de Física de la Materia Condensada, Universidad Autónoma de Madrid, Campus de Cantoblanco, E-28049 Madrid, Spain

[3] MIRE Group, Dpto. de Física de Materiales, Universidad Autónoma de Madrid, UAM, E-28049- Madrid, Spain

[4] Nano-Bio Spectroscopy Group and ETSF, Universidad del País Vasco, CFM CSIC-UPV/EHU, E-20018 San Sebastián, Spain

[5] Inst. Nicolas Cabrera, Univ. Autónoma de Madrid, E-28049 Madrid, Spain

[6] IKERBASQUE, Basque Foundation for Science, E-48013, Bilbao, Spain

[7] Instituto de Ciencia de Materiales de Madrid (ICMM-CSIC), Madrid, E-28049, Spain.

[†] Present address: Department of Physics, University of California, Santa Barbara CA 93106 USA

[‡] Present address: Institute of Photonics, Vienna University of Technology, Gußhausstraße 27-29, 1040 Vienna, Austria

[*] Present address: Instituto Madrileño de Estudios Avanzados en Energía (IMDEA-energía), Parque Tecnológico de Móstoles E-28935 Móstoles, Madrid, Spain

Email: jisland@physics.ucsb.edu, andres.castellanos@csic.es

**Abstract**   The isolation of graphene and transition metal dichalcongenides has opened a veritable world to a great number of layered materials which can be exfoliated, manipulated, and stacked or combined at will. With continued explorations expanding to include other layered materials with unique attributes, it is becoming clear that no one material will fill all the post-silicon era requirements. Here we review the properties and applications of layered, quasi-one dimensional transition metal trichalcogenides (TMTCs) as novel materials for next generation electronics and optoelectronics. The TMTCs present a unique chain-like structure which gives the materials their quasi-one dimensional properties such as high anisotropy ratios in conductivity and linear dichroism. The range of band gaps spanned by this class of materials (0.2 eV- 2 eV) makes them suitable for a wide variety of applications including field-effect transistors, infrared, visible and ultraviolet photodetectors, and unique applications related to their anisotropic properties which opens another degree of freedom in the development of next generation electronics. In this review we survey the historical development of these remarkable materials with an emphasis on the recent activity generated by the isolation and characterization of atomically thin titanium trisulfide ($TiS_3$).

# INTRODUCTION

Layered semiconductors are fast becoming strong candidates as next generation electronic and optoelectronic devices.[1-13] With the isolation of graphene by mechanical exfoliation[14] has come a tremendous research effort on discovering novel layered materials that can replace and outperform aging silicon devices but with the caveat of being easily integrated with current CMOS technologies. With each (re)discovery, atomically-thin layered materials often bring new functionalities and applications beyond those of industry standard transistors and photodetectors such as stronger light-matter interaction[15, 16], flexibility and transparency[17, 18], and the opportunity to combine layered materials into van der Waals heterostructures[19-21]. The most promising and studied layered semiconductors to date are the transition metal dichalcogenides ($MoS_2$, $MoSe_2$, $WS_2$, $WSe_2$). These materials, while holding many of the merits of layered materials, present a direct band gap only in single layer form due to interlayer interactions.[22, 23] Additionally, their bandgaps in single layer form range from ~1.6-2 eV, much larger than nanostructured silicon (~1.1 eV), making them unsuitable as a direct replacement.

The broad family of layered materials is only so far barely explored and there still exist many members that remain to be investigated.[24] The group IV-V transition metal trichalcogenides (TMTC), for instance, are interesting when compared with the well-studied dichalcogenides (TMDCs) because of their quasi-one dimensional properties stemming from a reduced in-plane structural symmetry. This character gives the TMTC class of materials strong anisotropies in their electrical and optical properties.[25-31] This anisotropy lends an additional degree of freedom in the fabrication of next generation electronics such as high mobility transistors benefiting from reduced backscattering from hot electrons[32-34] and novel integrated digital inverters[35]. More than this, these materials have great potential in electronics that rely on the generation and detection of polarized light, *i.e,* on-chip polarizers[36-39], polarization sensitive photodetectors[40-42], and polarized light emission[28, 43, 44].

Titanium trisulfide ($TiS_3$), in particular, has gained recent attention as it presents a robust direct band gap of ~1 eV[45,46,47] which varies little with layer thickness or stacking order.[46, 48] This robust quality is in direct contrast with the TMDC layered materials which vary greatly with thickness and strain[49-51]. Moreover, mobilities for $TiS_3$ transistors have been predicted to reach as high as 10000 $cm^2$/Vs for the high mobility axis[26] and simple transistors, with[52] and without optimizations[46], have been fabricated with reported mobilities as high as 70 $cm^2$/Vs. The range of band gaps spanned by the TMTCs also makes them well suited for photodetection. Photodetectors incorporating $TiS_3$ display high gains up to ~3000 A/W and fast switching times of a few milliseconds.[53]

The TMTCs first attracted attention in the beginning of the 1970s as viable materials for studies of electrical conduction in one dimension[54-56]. Spurred by theories from Peierls of a one-dimensional charge density wave (CDW) instability[57, 58] and by Fröhlich of superconductivity due to sliding of such a CDW[59], experimentalists set to conduct electrical measurements of the TMTC class of quasi-1D materials. Early studies revealed charge density wave formations in $NbSe_3$ at temperatures of 145 K and 59 K[60] and superconductivity with increased pressure[61].

Also during the 70s, it was shown that, like TMDCs, TMTCs can undergo a topotactic reaction with lithium making them viable materials as electrodes in lithium batteries.[62-64] There has been continued research in this direction[65, 66] and more recent applications of TMTCs include their use as precursors in the synthesis of two-dimensional (2D) materials (metal dichalcogenides and trioxides)[67, 68], anodes in photoelectrochemical cells[69-71], photoelectrodes for hydrogen photoassisted generation[72, 73], UV-visible photodetectors[53, 74-76] and lastly, in thermoelectric devices [77, 78].

In this review we discuss the unique properties of the group IV-V trichalcogenides and their use for applications in electronic devices and photodetection. In particular, we begin with a description of the crystal structure of the $MX_3$ class of materials. We then describe the synthesis and optoelectronic properties of the bulk material. We review the current technology for isolation of few-layer nanoribbon samples and report on characterizations of the samples showing that they largely retain the properties of the parent material, namely, high crystallinity and a range of electronic band gaps suitable for a variety of applications. The current state-of-the-art in electronic devices is reviewed with an emphasis on recently reported transistors and optoelectronic devices incorporating TMTCs and their characteristics reported in the literature. We finish this review with a look into the interesting research that is emerging on these materials resulting from their reduced in-plane structural symmetry which separates them from the more well-known TMDCs and other 2D materials in general.

# CRYSTAL STRUCTURE AND ELECTRONIC PROPERTIES

The $MX_3$ class of TMTCs are composed of transition metal atoms, M, belonging to either group IVB (Ti, Zr, Hf) or group VB (Nb, Ta) and chalcogen atoms, X, from group VIA (S, Se, Te), see Fig. 1(a). The molybdenum (Mo) and tungsten (W) (group VIB) type trichalcogenides have been synthesized but only amorphous semiconducting variants have been reported[79-86]. The basic structure for these materials consists of covalently bonded layers, such as those shown in Fig. 1(b) for $TiS_3$, which are bonded through van der Waals forces. A single layer is formed by $MX_3$ (Fig. 1(c)) chains which are built from $MX_6$, wedge-shaped triangular prisms along the *b* crystal axis. These chains are covalently bonded within the *a-b* plane (Fig. 1(c)).

Titanium trisulfide, in particular, crystallizes in the monoclinic $ZrSe_3$-type crystal structure. It belongs to space group $C_{2h}^2$ ($P2_1/m$) with two formula units per unit cell.[87] Ti atoms, at the center of each prism, are coordinated to six S atoms at the corners and another two S atoms in neighbor chains. The prisms (basic coordination units) are linked via shared faces into an infinite chain parallel to the crystallographic b-axis, the direction with the smallest Ti-Ti separation. Parallel neighboring chains are connected forming a sheet-like unit (parallel to the crystal face (001)). These sheets are bonded one to another by weak van der Waals bonds. Two of the three S atoms in the triangle of the prismatic coordination are bonded in a single disulfide ion ($S_2^{2-}$) and the third is formally sulfide ($S^{2-}$). Two layers of sulfur dimers form the van der Waals gap (parallel to the crystallographic c-axis). The bond axis of the disulfide group is parallel to the a-axis.

The M-M and X-X bond lengths have a strong influence on the electrical and optical properties of the TMTCs.[88, 89] $ZrSe_3$, for instance, is semiconducting due to the strong chalcogen-chalcogen bonds which retain two electrons per unit.[90] The group IV TMTCs crystalize in a monoclinic structure and are mainly semiconducting, excluding the telluride varieties which are small band gap semiconductors. Table 1 shows an overview of the electronic properties of the group IV and group V TMTCs. The group V TMTCs contain materials that present collective electronic phases such as charge density wave instabilities and superconductivity. $NbS_3$ (triclinic structure) presents a resistive anomaly at 340 K due to CDW formation. While $NbSe_3$ (monoclinic structure) presents CDW induced resistivity anomalies at 145 K and 59 K. $TaS_3$ (orthorhombic structure) also shows CDW formation but $TaSe_3$ presents a superconducting transition at a temperature of 2.1 K.

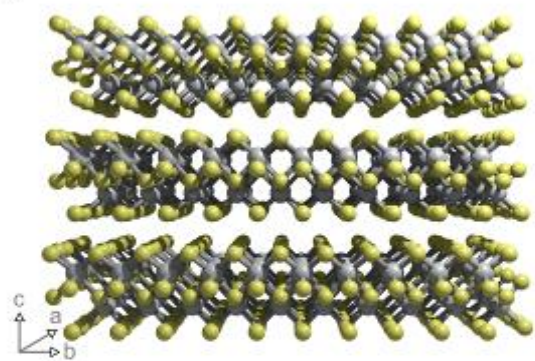
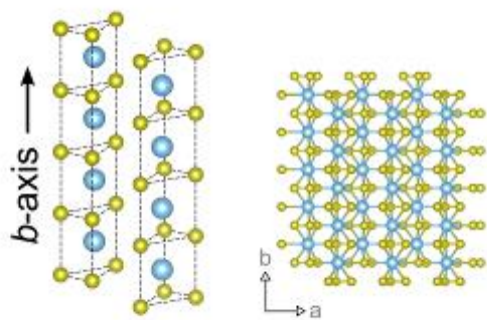
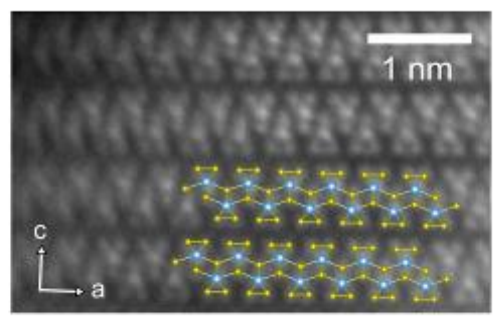
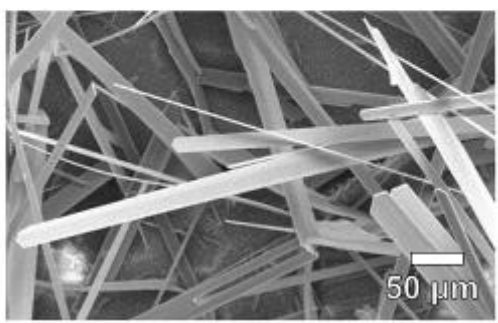
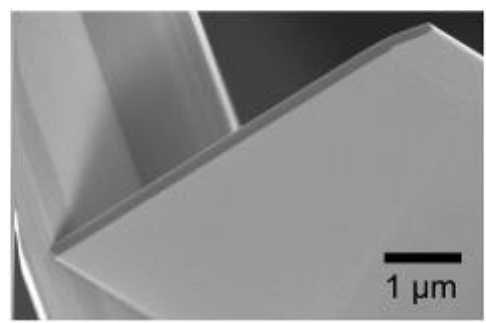

Figure 1. (a) Elements that make up the group IV-V TMTCs. (b) Three dimensional model of the crystal structure of $TiS_3$. This panel is adapted from ref. [46] with permission. (c) b-axis chains that lead to the quasi-one dimensional properties of the TMTCs. a-b plane of the crystal structure showing the covalent bonding of the b-axis chains in a single layer of $TiS_3$. (d) TEM of the cross-section of a $TiS_3$ sample showing the layered structure and the ends of the b-axis chains. Reprinted (adapted) with permission from [77]. Copyright 2014 American Chemical Society. (e) Scanning electron microscopy image of a film of $TiS_3$ nanoribbons grown at 550ºC (1 hour). This panel is adapted from ref. [91] with permission from the author. (f) Scanning electron microscopy image of a single $TiS_3$ nanoribbon. This panel adapted from [53].

Table 1. Overview of the electronic properties of the TMTCs

| Group | Transition Metal | Chalcogen | | |
|---|---|---|---|---|
| | | S | Se | Te |
| IV | Ti | **Monoclinic** Diamagnetic Semiconductor (0.8-1 eV)[45, 92, 93] | No reported synthesis | No reported synthesis |
| | Zr | **Monoclinic** Diamagnetic Semiconductor (1.9-2.1 eV)[92, 93] | **Monoclinic** Diamagnetic Semiconductor (1.1-1.3 eV)[92, 93] | **Monoclinic** Superconducting (2K)[94] |
| | Hf | **Monoclinic** Diamagnetic Semiconductor (1.95 eV)[92] | **Monoclinic** Diamagnetic Semiconductor (1.02 eV)[92] | **Monoclinic** Small gap semiconductor (~0.2 eV)[92] |
| V | Nb | **Triclinic** Diamagnetic Semiconductor, (0.83 eV)[95] CDW (340 K)[96] | **Monoclinic** Metal, CDW (145 K) (59 K)[97] | No reported synthesis |
| | Ta | **Orthorhombic** Metal, CDW (210-220 K)[98-100] | **Monoclinic** Metal, Superconducting (2.1 K)[101] | No reported synthesis |

# SYNTHESIS AND BULK CHARACTERIZATION

The synthesis of the TMTCs started to be investigated in the sixties of the past century due to the great variety of possible applications associated with their wide range of physical properties[87, 93, 102, 103].

One of the most interesting materials of the family is $TiS_3$ due to its low toxicity and the abundance and moderate cost of its constituent elements. Since the first synthesis of $TiS_3$ by Blitz[104] the most used growth process has been direct sulfuration of titanium powder in vacuum with sulfur powder in excess. Usually the synthesis is carried out under a temperature gradient and the material grows in the cold extreme of the ampule. The sulfuration temperature and gradient used vary with each study but the most frequently used values are around 500ºC and 50-100ºC, respectively[91]. This method has also been widely used to synthesize other metal trisulfides[93, 103, 105, 71, 73, 75, 76, 106, 107]. It is commonly included in the group of chemical vapor transport (CVT) techniques where sulfur or $TiS_x$ species are considered as transport agents. However, properly said CVT refers to synthesis with transport agent species different from the reactive ones, as with halogens, which have also been utilized to grow some TMTC compounds[76, 105, 108]. Detailed reviews on the TMTC single crystal growth by chemical vapour transport were reported by Srivastava[109] and Levy[110]. Less often used is the growth process based on metalorganic sulfur precursors in metalorganic chemical vapor deposition (MOCVD)[111-113]. The above mentioned methods are used to grow single crystals and bulk powders, however, densified powder and pellets have been obtained under pressure[114, 115] and by spark plasma sintering[77, 78]. In addition, the as grown TMTC powders have been dispersed in a colloidal suspension and deposited as films by drop casting on different substrates[73, 75]. Finally, $TiS_3$ nanobelts have been grown as well from the sulfuration of Ti sheets or disks[70-72, 116] and previously deposited Ti thin films[45, 68]. Table 2 summarizes the different reported methods (in different colors) used for the synthesis of TMTC compounds. Earlier reviews of synthesis methods for these materials can also be found in refs. [109, 110]

Table 2. Growth processes used in the synthesis of $MX_3$ (TMTC) compounds.

| Compounds | Form | Method | Refs. |
|---|---|---|---|
| $TiS_3$, $HfS_3$, $ZrS_3$, $NbS_3$, $TaS_3$ | Single-crystals, powders, pellets | Metal+S solid-gas direct reaction | 45, 69, 71, 103-107, 117-125 |
| $ZrS_3$ | Exfoliated nanosheets | Zr +S solid -gas reaction Liquid-phase exfoliation | 71 |
| $TiS_3$ | Exfoliated nano-ribbons and nanosheets | Ti + S solid-gas reaction, Mechanical cleavage | 27, 46, 53 |
| $HfS_3$ | Exfoliated nano-belts | Hf + S solid-gas reaction, Liquid phase exfoliation | 75 |
| $ZrS_3$ | Exfoliated nano-belts | Ti + S solid-gas reaction, Mechanical cleavage | 76 |
| $TiS_3$, $ZrS_3$, $ZrSe_3$, $ZrTe_3$, $HfS_3$, $HfSe_3$, $HfTe_3$ | Single crystals | Metal + Chalcogen chemical vapor transport with iodine as agent transport | 87, 92, 105, 108, 126-129 |
| $TiS_3$, $ZrS_3$, and ternaries | Single crystals | Ti, Zr + S chemical vapor transport with bromine as agent transport | 130, 131 |
| $Ti_3S_4$, $TiS_2$, $TiS_3$ $ZrS_3$ | Powders and thin films | $MCl_4$ + organic sulfurizing agents MOCVD[i] | 112, 113, 129 |
| $TiS_3$ | Nanobelt films, Nanoribbon films | Ti foils, disks, powders, films + S direct reaction | 45, 70, 72, 116 |
| $TiS_3$, $TiS_2$ | Films | Ti film +Ti powder +S SACVT[ii] | 68 |
| $TiS_3$, $ZrS_3$, $HfS_3$ | Films by drop casting | Metal powder + S, drop casting | 73 |
| $NbS_3$, $NbSe_3$, $TaS_3$, $TaSe_3$, | Powder, pellets | Metal + S direct reaction at 2GPa | 114, 115 |
| $TiS_3$, $Ti_{1-x}Nb_xS_3$ | Powder, pellets | Ti powder + S direct reaction +SPS[iii](50-85MPa) | 77, 78 |

[i]MOCVD, Metal Organic Chemical Vapor Deposition
[ii]SACVT, Surface Assisted Chemical Vapour Transport
[iii] SPS, Spark Plasma Sintering

It is known that the growth of nanowires, nanobelts, and other nanostructure morphologies may be explained by specific growth mechanisms such as vapour-liquid-solid (VLS), solution-liquid-solid (SLS) etc.[132]. It has been recently reported[91] that the growth mechanism of $TiS_3$ nanobelts takes place by means of a gaseous species, i.e. by a vapour-solid growth mechanism (VS), as occurs with other semiconducting materials[133]. The VS mechanism of crystal growth was already purposed and demonstrated by Burton et al[134]. Other authors[68, 69] have also suggested this mechanism in the synthesis of TiS$_3$ but without evidence about the existence of any gaseous species.

One evidence of this type of growth (VS) is the pointed shape of the end of the nanobelts, as can be clearly observed in the SEM image of Figure 1(e), which appears independent of the sulfuration temperature used in the synthesis process. In this mechanism the preferential growth direction is determined by the competition between the growth rates of the different crystallographic planes. The planes with the highest surface free energy are the most reactive (less stable) and the preferential growth will take place on that surface. TMTC powders can be easily deposited on different substrates by drop coating to be used in form of thin films[73]. This method has been successfully used to deposit MS$_3$ (M=Ti, Zr, Hf, Nb) on different substrates. Figure 2 show the images of these films beside their SEM micrographs.

In this synthesis, powders of TiS$_3$, ZrS$_3$, HfS$_3$, and NbS$_3$ have been obtained by direct sulfuration of metal (Ti, Z, Hf and Nb) in a vacuum (P < 10$^{-2}$ mbar) sealed ampoule filled with sulfur powder (~mg) at 550 ºC during 60 hours. Sulfuration temperatures higher than 600 ºC-650ºC lead to the decomposition of MS$_3$ forming the MS$_2$ phase. A heating rate of 0.4 ± 0.05 ºC/min is used and the cooling process takes place in ambient conditions. During the synthesis, the sulfur pressure in the ampoule is determined by the liquid-gas equilibrium vapour pressure of sulfur at a given temperature and reaches values between 10$^{-1}$ bar to 10 bar. At the growth temperatures, sulfur gas is mainly formed by the biggest species (S$_8$, S$_7$ and S$_6$). After the synthesis, dense films can be prepared by dispersing the powder in an organic solvent to produce a colloidal suspension and drop casting deposition technique. Figure 2 shows optical and SEM images of films resulting from this method for four members of the MX$_3$ family of crystals.

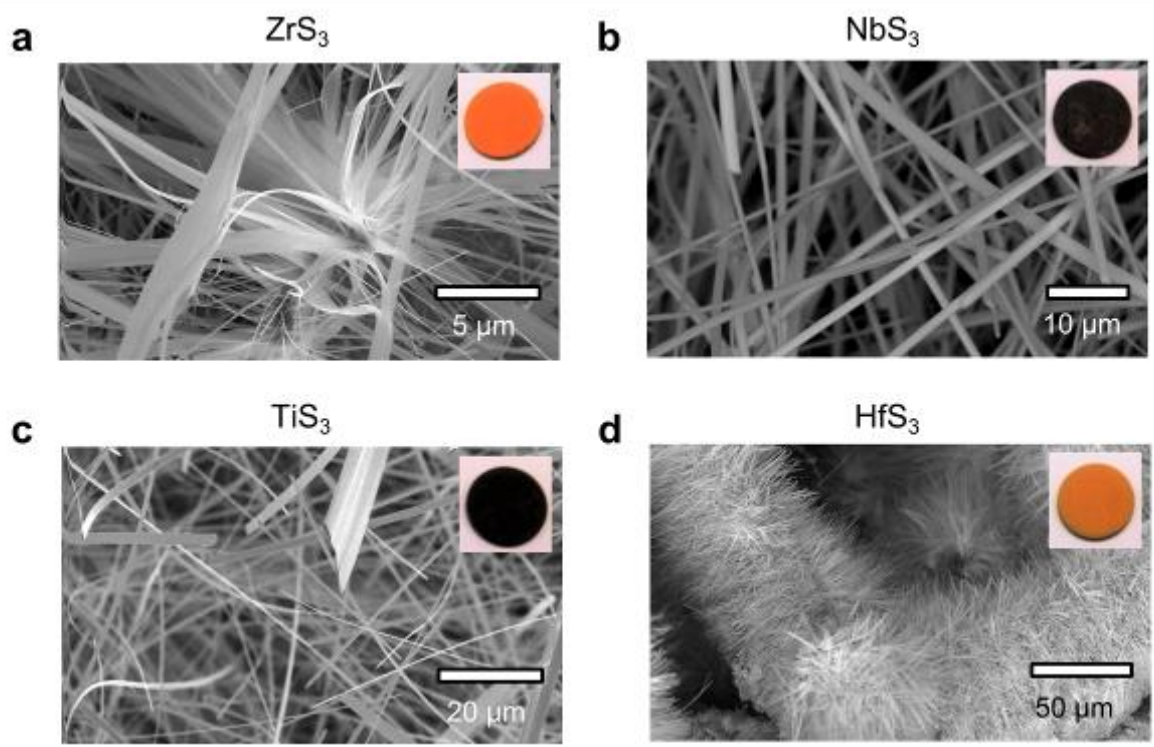

Figure 2. (a) SEM image of ZrS$_3$ film deposited on a Ti substrate by drop casting of a colloidal suspension. (b) SEM image of NbS$_3$ film deposited on a Ti substrate by drop casting of a colloidal suspension. (c) SEM image of a TiS$_3$ film deposited on a Ti substrate by drop casting of a colloidal suspension. (d) SEM image of HfS$_3$ film deposited on a Ti substrate by drop casting of a colloidal suspension. The insets for all panels show an optical image of the film on the Ti substrate with a diameter of a few millimetres. This figure comprises unpublished work.

## *Optical absorption in bulk material*

The optical energy gap of bulk MX$_3$ material has been determined from optical absorption measurements. Optical absorption of TiS$_3$ has been investigated from diffuse reflectance measurements of TiS$_3$ powders[45] by using the Kubelka–Munk function (F(R)), defined by F(R)=(1-R)$^2$/2R, where R is the diffuse reflectance. This function is commonly used as an approximation of the optical absorption coefficient.[135] There, it was used to obtain a Tauc plot (F(R)·hν)$^2$ *vs.* hν to determine the value of the direct energy gap of TiS$_3$. From a linear fit of these results two direct transitions of 0.94 ± 0.02 eV and 1.18 ± 0.02 eV are obtained. In addition, it was also estimated by photoelectrochemical measurements. The estimated band gap (1.07 eV) obtained from the photocurrent density spectral response[53] was in good agreement with those from optical measurements.

Recently, energy band gaps of bulk MS$_3$ (M=Ti, Zr, Hf) have been obtained by measuring the optical absorption response (reflectance and transmittance) of drop coated films deposited on quartz from their respective powders[73]. Figure 3 shows the corresponding Tauc plots of the absorption coefficient used to determine the energy band gap of TiS$_3$, ZrS$_3$, and HfS$_3$ beside that of NbS$_3$ films also deposited by drop casting (from NbS$_3$ powder), including their values, which agree well with those previously reported and listed in Table 1.

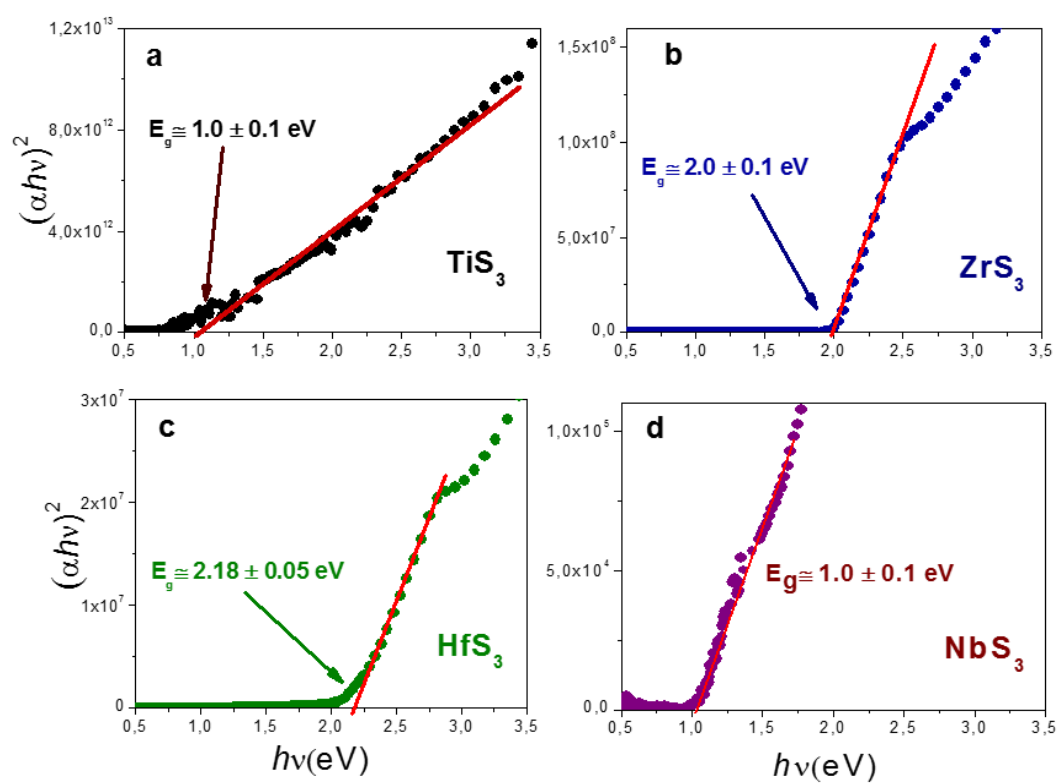

Figure 3. Tauc plot of the absorption coefficient for different MX$_3$ films deposited by drop casting from their respective powders. Panels (a-c) are adapted from ref. [73] with permission. Panel (d) is unpublished data.

### *X-ray Diffraction (XRD) Characterization of the Bulk Material*

X-ray diffraction patterns of TiS$_3$, ZrS$_3$, HfS$_3$ and NbS$_3$ films (Fig. 4) have been reported and are consistent with JCPDS 015-0783, JCPDS 030-1498, JCPDS 029-0655 and JCPDS 071-0468 for TiS$_3$, ZrS$_3$, HfS$_3$ and NbS$_3$, respectively. The most intense XRD peak corresponds to the (001) plane. Crystallite sizes of the different TMTC are estimated from the half height width of the (001) XRD diffraction peak, according to the Scherrer formula.[136] Values obtained are shown in table 3 besides the stoichiometric S/M ratios estimated by EDX.

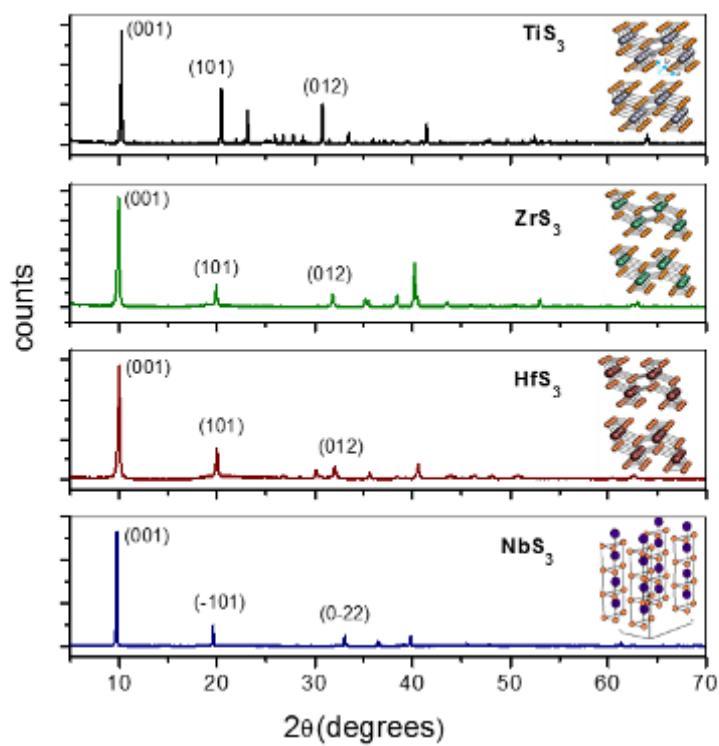

Figure 4. X-ray diffraction patterns of monoclinic a) TiS$_3$, b) ZrS$_3$, c) HfS$_3$ and anorthic d) NbS$_3$ samples. Inset images correspond to the crystal structure. Panels (a-c) are adapted from ref. [73] with permission. Panel (d) is unpublished data

Table 3. MX$_3$ crystalline and compositional characteristics (Data of TiS$_3$, ZrS$_3$ and HfS$_3$, are reported in ref. [73])

| Material | Crystalline size (nm) | EDX Stoichiometic ratio S/M | |
|---|---|---|---|
| | | powder | film |
| TiS$_3$ | 79 ± 2 | 2.97 ± 0.05 | 2.90 ± 0.05 |
| ZrS$_3$ | 37± 2 | 2.85 ± 0.05 | 2.70 ± 0.05 |
| HfS$_3$ | 34± 2 | 2.93 ± 0.05 | 2.85 ± 0.05 |
| NbS$_3$ | 120± 5 | 2.94 ± 0.05 | 2.72 ± 0.05 |

## ISOLATION AND EXFOLIATION OF MX$_3$ CRYSTALS

The isolation of individual whiskers of MX$_3$ crystals for bulk studies has been achieved previously in several studies.[120, 121, 124, 125, 137-139] These whiskers typically had thicknesses of a few to several micrometres. Contact was made by gold filled epoxy or by mechanically depressing indium contacts onto individual whiskers. Mechanical exfoliation, using sticky tape or viscoelastic materials, has allowed exfoliation of individual nanoribbons down to a few layers.[53, 140] The high aspect ratio of individual nanoribbons makes it challenging to isolate single layers. Recent calculations show that the interlayer potential (45 meV/atom), is less than that of MoS$_2$ (60 meV/atom) and therefore should allow for exfoliation of the bulk down to a single layer.[46] Indeed, the 10 nm thickness barrier was recently overcome through control of the morphology of the bulk material to produce flakes instead of ribbons. The flakes, with lower aspect ratio, allow exfoliation down to a single layer.[46]

Direct mechanical exfoliation is achieved by lightly placing a polydimethylsiloxane (PDMS) stamp on a bulk growth disk so as not to pick up too much material (see Figure 5(a)).[46] A high density of nanoribbons makes it difficult to later pattern metal contacts to individual ribbons. The PDMS stamp is then let to adhere to a Si/SiO$_2$ substrate and subsequently slowly lifted from the substrate, isolating and at the same time exfoliating individual nanoribbons (Figure 5(a)). Figure 5(b) shows a transmission mode optical image of a PDMS stamp with TiS$_3$ material. Figure 5(c) shows an optical image of TiS$_3$ nanoribbons on the surface of a SiO$_2$ substrate after transfer. As with most layered materials, the thinner material is more transparent and shows a purple/blue color on 285 nm of SiO$_2$ (Figure 5(d-g)) while the thicker ribbons are more opaque and have a yellow color.

Liquid phase exfoliation of MX$_3$ has also been recently achieved by Stolyarov et al. for TaSe$_3$ using ethanol.[141] The method starts with 6 mg of bulk powdered TaSe$_3$ which is sonicated in 10 ml of ethanol for several hours. The mixture is then centrifuged at 2600 rpm for 15 min to remove large particles. This method produced nanoribbons with widths of approximately 30 to 80 nm which were then subsequently exfoliated further by mechanical means. Further colloidal dispersions have been achieved for TaS$_3$ in MeCN and DMSO[142], and for NbS$_3$ and NbSe$_3$.[143, 144] Figure 5(h) shows the Tyndall effect for a colloidal dispersion prepared by Fedorov et al. of NbSe3 in acetonitrile. Chemically exfoliated NbS$_3$ nanoribbons, using the same preparation method, are shown in Fig. 5(i) in an AFM image. A closer look at a single ribbon, Fig. 5(j), shows a thickness of a few nanometers (inset).

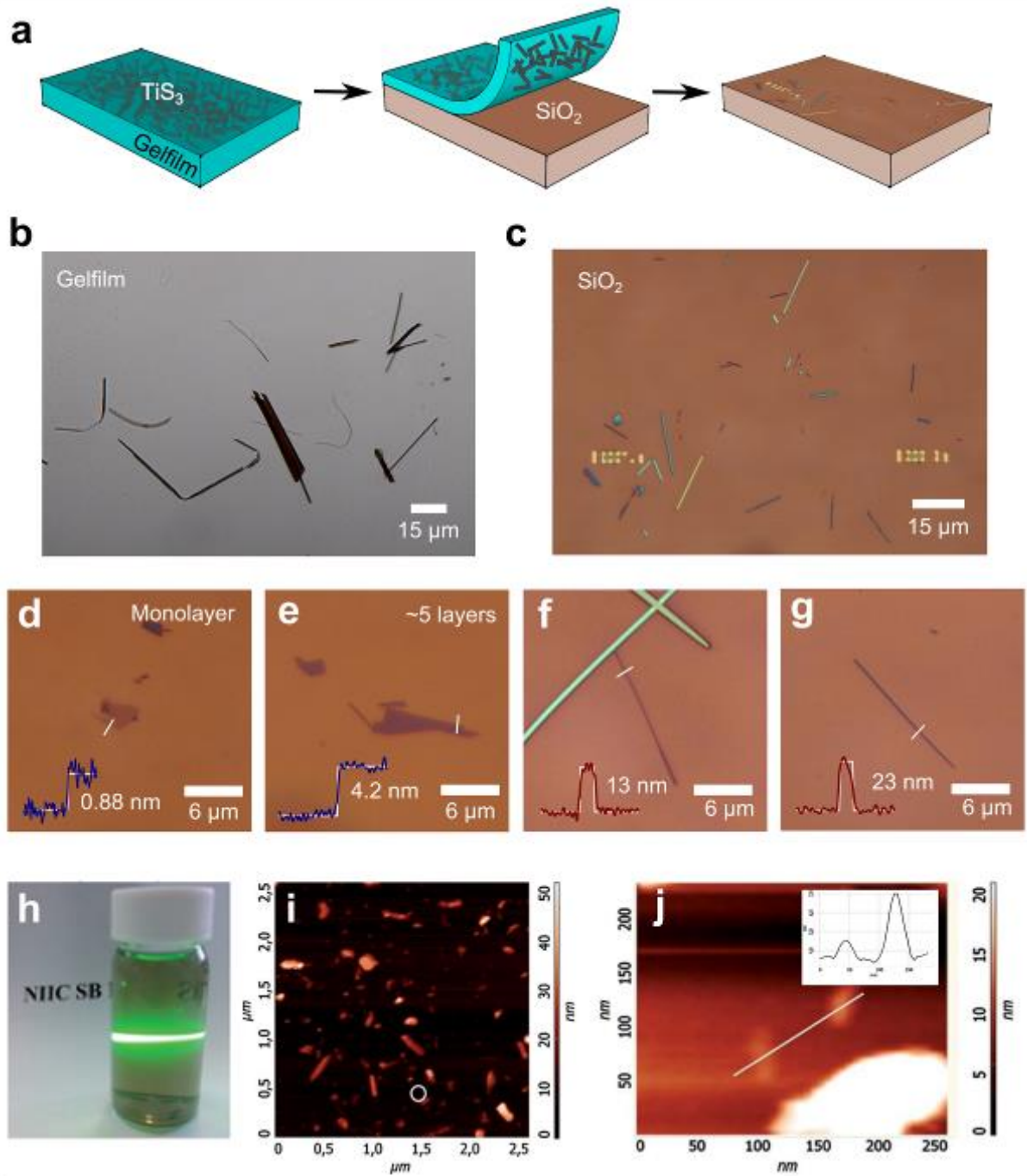

Figure 5. (a) Viscoelastic stamp with $TiS_3$ nanoribbons on the surface. The stamp is placed on a receiving (Si/SiO$_2$) substrate and peeled off leaving behind isolated and exfoliated nanoribbons. This Panel adapted from ref. [47] (b) Transmission mode optical image of $TiS_3$ nanoribbons on PDMS. (c) Optical image of $TiS_3$ nanoribbons on Si/SiO$_2$. Panels (b-c) are unpublished images. Optical images of monolayer $TiS_3$ (d), few layer (~5) $TiS_3$ (e), a 13 nm $TiS_3$ nanoribbon (f), and a 23 nm $TiS_3$ nanoribbon (g). Panels (d-g) are adapted from ref. [46] with permission. (h) The Tyndall effect for a colloid dispersion of $NbSe_3$ in acetonitrile. (i) AFM image of exfoliated $NbS_3$ nanoribbons. (j) AFM image of two individual $NbS_3$ nanoribbons with thicknesses of a few manometers. Panels (h-i) are adapted from ref. [143] with permission.

# ELECTRONIC STRUCTURE

The electronic band structure is closely related to the electronic and optical properties of a material. Hence, it is an essential tool for understanding, predicting, and fine-tuning the optical and electronic response of a system. Recently, a lot of attention has been drawn to the theoretical understanding of the electronic band structure of layered TMTCs within density functional theory (DFT) and many-body methods. For the bulk material it has been theoretically found that $TiTe_3$, $HfS_3$, $HfSe_3$, $ZrS_3$, and $ZrSe_3$ are indirect band gap semiconductors with band gaps in the range of 0.44 to 2.04 eV[145]. At the same time, bulk $TiSe_3$, $ZrTe_3$, and $HfTe_3$ are predicted to have the electronic structure of metals [146, 147]. More precisely the band gap of bulk $ZrS_3$, $ZrSe_3$, $HfS_3$, $HfSe_3$, and $NbS_3$ are 1.87 eV, 0.68 eV, 1.90 eV, 0.59 eV, and 1.07 eV, respectively [147], which is in reasonably good agreement with experiments[93, 148-152] (see Table 1). For example, the measured band gap of $ZrS_3$ and $HfS_3$ is 1.91 and 1.95 eV, respectively. In addition to the bulk material, some monolayer TMTCs have been theoretically modeled within a hybrid functional approach to DFT [25, 147]. $ZrS_3$, $ZrSe_3$, $HfS_3$, $HfSe_3$, and $NbS_3$ monolayers are found to be semiconductors with an indirect band gap of 1.92, 0.92, 1.94, 0.80 and 1.18 eV, respectively [25]. Furthermore, monolayer $ZrTe_3$ and $HfTe_3$ are predicted to have the electronic structure of metals [25].

Theoretical investigations show that all TMTCs except $TiS_3$ are either metals or semiconductors with an indirect band gap. However, calculations also predict that monolayer and few-layer $TiS_3$ are semiconductors with a direct gap [65, 153, 154]. DFT calculations with the Perdew-Burke-Ernzerhof's functional (DFT-PBE) predict a direct gap of monolayer $TiS_3$ at about one third of its experimental value, while hybrid functional HSE06 [65], or a non-self-consistent $GW$ method ($G_0W_0$) are able to bring the theoretical value of the gap in closer agreement to the experimental value. In Figure 6(a) the band structure of a monolayer of $TiS_3$ is shown as calculated via DFT-PBE and $G_0W_0$ in the plasmon pole approximation: while DFT underestimates the direct gap at the $\Gamma$ point, $G_0W_0$ opens up the gap to a value of around 1.14 eV (see Figure 6(a)), by shifting the valence and conductance bands apart. This is in good agreement with experimental values for thin films, where a direct optical transition at 1.10 eV has been detected [47] and to previous measurements [46]. Additionally, in Figure 6(b) one can see that the band gap is almost constant independently of the number of layers, showing the small influence the van der Waals force has on the electronics properties of $TiS_3$. In Figure 6(c), for comparison, we report the band structure of a few selected monolayer TMTCs calculated with the hybrid functional HSE06 [25]. We see a quite large variability of the band structure and electronic gap, suggesting that these materials can satisfy a diverse range of technological requirements.

The difference between the optical and the electronic gap gives rise to an excited state with a certain binding energy, an electron-hole exciton. Experimentally, a binding energy of the order of 130 meV has been measured for $TiS_3$ ribbons. This value is in good agreement with Bethe–Salpeter equation calculations (100 meV) (see the Optoelectronic Properties and Devices section for more details). [47] All this shows that, on the one hand, state-of-the-art many-body theory calculations are capable of describing the optical and electronic properties of this layered material within high accuracy. One the other hand, calculations on bulk $TiS_3$ indicate that the material has an indirect band gap [65, 153] in discrepancy with experiments. The indirect nature of the gap was obtained within DFT-PBE [153], hybrid functional [65], and also $GW$ calculations in the plasmon pole approximation. It is believed that this discrepancy between theory and experiment is still an open question and further investigation is necessary, also to validate the predictive power of the theory. In particular, it would be interesting to see if the band structure for bulk $TiS_3$ can be experimentally determined, outside of the center of the first Brillouin zone.

Insights into the relation between the band and the crystal structure open up the possibility of tuning the optical and electrical properties as desired. One possibility is to introduce vacancies in or allow for oxidation of the pristine crystal structure of $TiS_3$. [46, 153] It has been found that the presence of certain S vacancies creates a (fully occupied) localized electronic state below the Fermi energy, which increases the free carrier (electron) density. [46] This n-type doping decreases the resistivity and increases the mobility. In contrast, for the monolayer, superficial S vacancies lightly affect the electronic properties. [153] Recently, considerable attention on strain induced band-gap modulation on $MX_3$ has been drawn[25, 26, 48, 147, 154-157]. The ability to control the band gap and its nature can have a wide impact on the use of TMTCs, and in particular $TiS_3$, for optical applications. Indeed, by inducing tensile or compressive strain in certain directions of the unit cell one can either increase or decrease the band gap [155]. In addition to this, a direct-indirect gap transition can be induced in TiS$_3$ layered materials by applying a compressive strain in the easy electronic transport direction [155]. This opens up the door of experimentally engineering the electronic and optical properties of the material, separately.

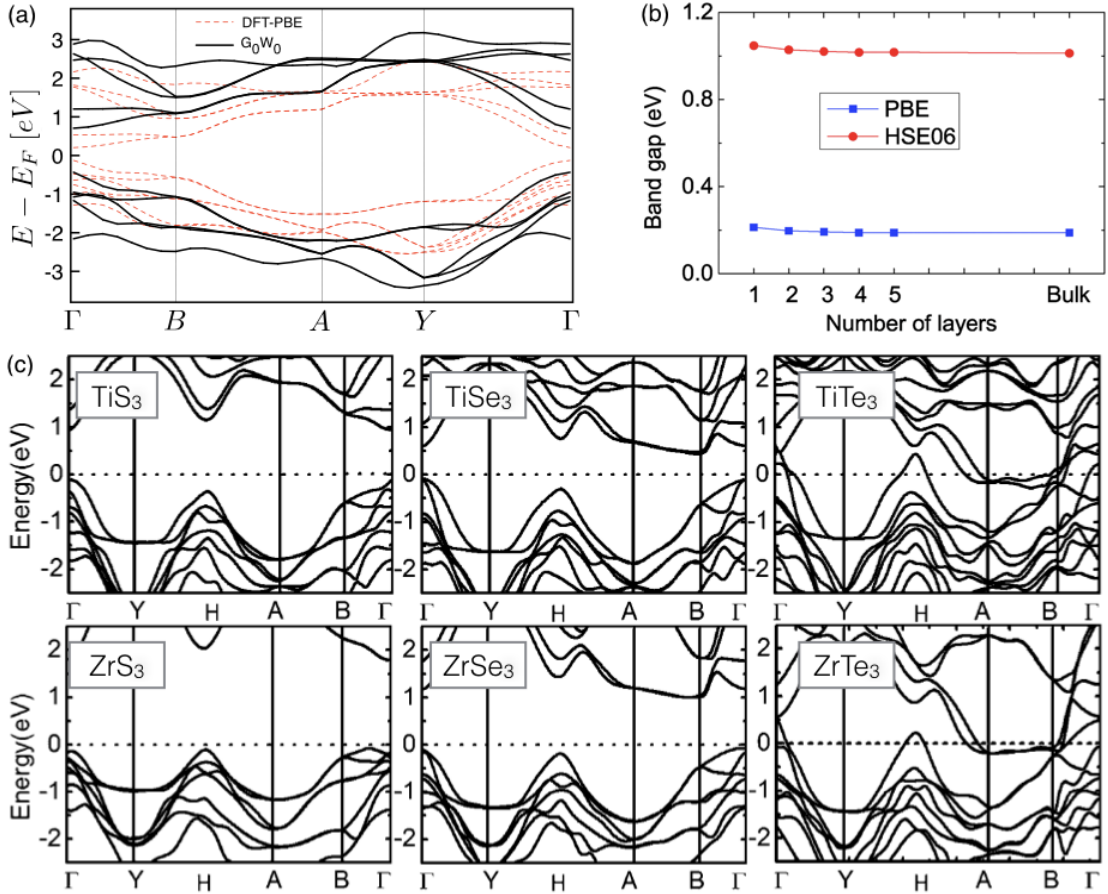

Figure 6. (a) The electronic band structure for monolayer TiS$_3$ obtained with DFT and $G_0W_0$. DFT underestimates the band gap by about a factor 1/3 with respect to both the $G_0W_0$ and the experimental values. This panel is comprised of unpublished work. (b) Band gap for TiS$_3$ as a function of thickness. This panel is adapted from ref. [48] with permission. (c) Electronic band structure for some selected monolayer TMTCs using the HSE06 functional. This panel is adapted from ref. [25] with permission. For (a) and (c), the high-symmetry points in the two-dimensional reciprocal space are $\Gamma = (0,0)$, $Y = (0,1/2)$, $A = (1/2,1/2)$, $B = (1/2,0)$.

## ELECTRICAL PROPERTIES AND DEVICES

Characterizations of electrical devices employing exfoliated individual MX$_3$ nanoribbons has been recently performed by several groups.[52, 53, 141, 158] The first basic devices were fabricated using standard electron-beam lithography procedures to create gold electrodes contacted to individual, exfoliated nanoribbons.[53] Figure 7(a) shows an atomic force microscopy (AFM) image of a representative device from the study. Five gold contacts are connected to and individual nanoribbon with a thickness of 28 nm. Figure 7(b) shows the measured drain current as a function of bias voltage for varying back-gate voltages and Figure 7(c) shows the field-effect transistor (FET) transfer curves for varying source-drain bias voltages. Of the 10 devices measured in the study, the highest mobility reported was 2.6 cm$^2$/Vs. Larger mobilities were subsequently achieved in FETs employing TiS$_3$ nanosheets grown at a different temperature.[46] In these devices, mobilities reached as high as 73 cm$^2$/Vs.

Shortly after, Lipatov et al. reported similar TiS$_3$ nanoribbon devices with enhanced performance by employing an atomic layer deposition (ALD) deposited layer of aluminum oxide.[52] Figure 7(d) shows a device schematic for the study. Chromium (3 nm) and gold (20 nm) electrodes were fabricated on top of individual, mechanically exfoliated nanoribbons on Si/SiO$_2$ substrates. Linear regime mobilites for as-fabricated devices reached as high as 18-23 cm$^2$/Vs and ON/OFF ratios were reported with a range of 30-300. Figure 7(e) shows the measured current as a function of back-gate voltage for one device. The linear regime mobility is estimated to be 20.3 cm$^2$/Vs. Figure 7(f) shows the same data on a log scale. In order to enhance the mobility through dielectric screening, a 30 nm layer of Al$_2$O$_3$ was deposited on the same device and measured again. Figure 7(e) shows the transconductance for the same device after ALD deposition showing an increase in the mobility to 36.8 cm$^2$/Vs. Figure 7(f) shows the same data on a logarithmic scale where it can be seen that the OFF current is more than an order of magnitude lower than the OFF current for the device without a dielectric screening layer. This translates to an improved ON/OFF ratio from 300 to 7100.

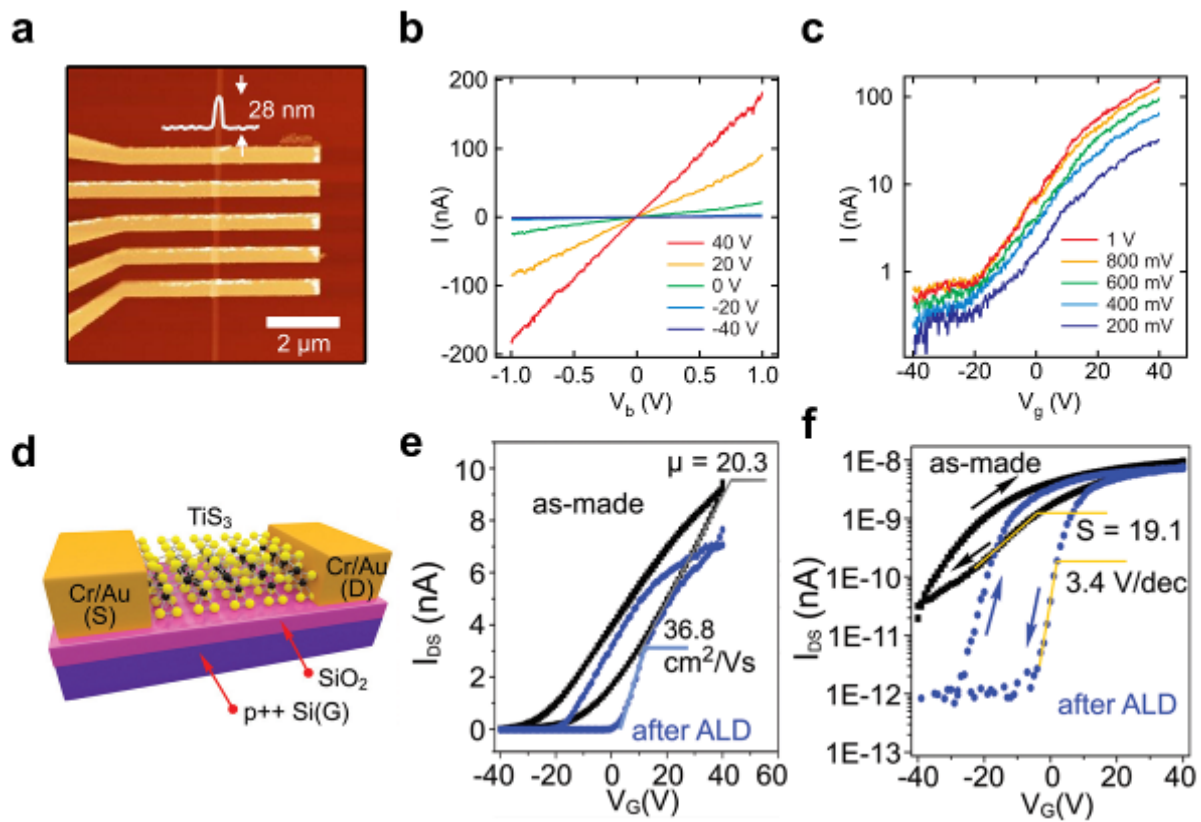

Figure 7. (a) AFM scan of a TiS$_3$ nanoribbon transistor. The nanoribbon has a thickness of 27 nm. (b) Current-voltage characteristics of the TiS$_3$ transistor pictured in panel (a). (c) FET transfer curves for the device pictured in panel (a). Panels (a-c) adapted from Ref. [46] with permission. (d) Model representation of an individual TiS$_3$ nanoribbon transistor. (e) Transfer curves for a TiS$_3$ nanoribbon transistor without (black curve) and with (blue curve) an ALD deposited Al$_2$O$_3$ dielectric layer. (f) Same data from (e) plotted on a log scale. Panels (d-e) adapted from Ref. [52] – published by The Royal Society of Chemistry.

Recently, Stolyarov et al. reported on the breakdown current density of TaSe$_3$ nanowires with a hexagonal boron nitride capping flake for use as electrical interconnects.[141] The boron nitride capping layer was used to facilitate in heat dissipation and to protect the nanowire from oxidation. Figure 8(a) shows a model representation of the devices studied. A single exfoliated TaSe$_3$ nanowire on Si/SiO$_2$ substrate is covered with a hexagonal boron nitride flake. Electrodes are patterned into the device by first etching the boron nitride with $SF_6$ and then depositing gold electrodes. In the study, several different sticking layers (Cr, Ti, Au, Pd) were attempted in order to make better contact to the nanowire but no major differences were found between the metals. Figure 8(c) shows the high field current-voltage characteristics for one device. The breakdown current density reaches 32 MA/cm$^2$ which is 18 times larger than state-of-the-art Cu interconnect technology. Using pulsed measurements, the authors showed that the primary breakdown mechanism is due to electromigration (primarily surface diffusion) and not self-heating which typically occurs in the carbon based nanomaterials such as graphene and carbon nanotubes. In a follow-up work, Liu *et al*. have shown that these devices also present low 1/f noise when compared to other low-dimensional materials like graphene and carbon nanotubes.[158]

Metallic nanowires that can stand high current densities like $TaSe_3$ are attractive for several applications, however, electronic applications require semiconductors, instead of metals, that can also carry high current densities before breakdown. It has been reported by Molina-Mendoza *et al*. that single $TiS_3$ nanoribbons can stand current densities as high as 1.7 MA/cm$^2$ when integrated in field-effect transistors, which is among the highest reported for semiconducting nanomaterials[159]. The $TiS_3$ nanoribbon field-effect transistors studied consist of single $TiS_3$ nanoribbons transferred onto a Si/SiO$_2$ substrate and contacted with Ti/Au electrodes by standard electron-beam lithography. By increasing the drain-source voltage and measuring the current passing through the nanoribbon, the authors report electrical breakdown of the devices and determine the current density just before failure (Figure 8(d)). An AFM image of the device after breakdown is shown in Figure 8(e), where features related to the material degradation can be identified. The analytical solution to the 1D heat equation is used to study the thermal mechanisms responsible for the electrical breakdown of $TiS_3$ nanoribbons, suggesting that the presence or creation of defects in the material are responsible for the breakdown in the devices with an estimated temperature of ~ 400 °C. These results are in agreement with the degradation of $TiS_3$ to $TiS_{3-x}$ observed by thermogravimetric coupled to mass spectrometry analysis in bulk $TiS_3$ and density functional theory combined with Kinetic Monte Carlo simulations, which suggest that the electrical breakdown occurs due to the O-mediated desorption of S atoms and the subsequent creation of vacancies in the material. The high current density measured in $TiS_3$ nanoribbons make this material a new and interesting candidate for high power electronics.

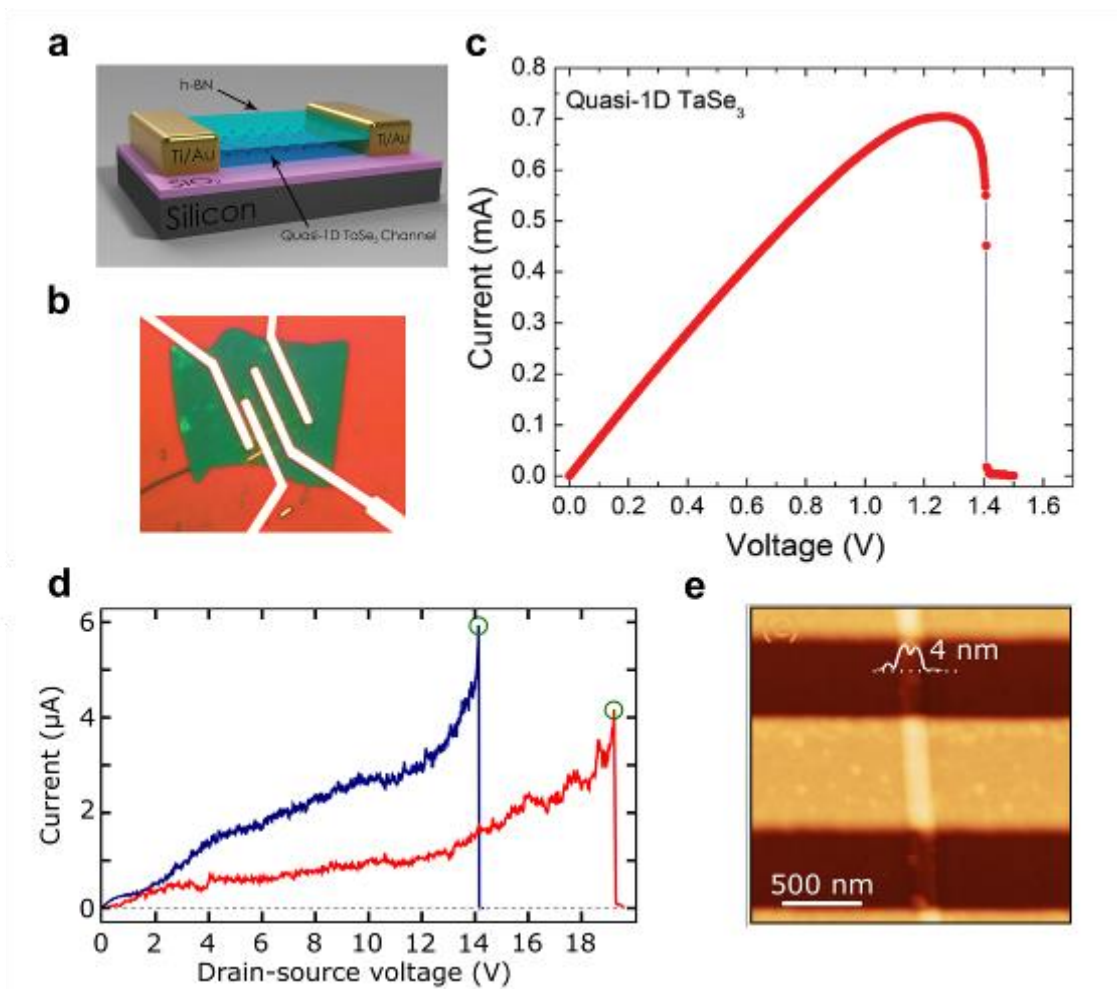

Figure 8. (a) Model representation of an individual TaSe$_3$ nanowire transistor with a hexagonal boron nitride flake used as a capping layer. (b) Optical image of an individual TaSe3 nanowire device. (c) Current-voltage curve of the breakdown characteristics of a TaSe$_3$ device. Panels (a-c) are adapted from Ref. [141]. (d) Current-voltage characteristics for two TiS$_3$ nanoribbon transistors, the green circles highlight the current-voltage value just before breakdown. (e) AFM of a TiS$_3$ nanoribbon transistor after electrical breakdown. Panels (d-e) adapted from ref. [159] with permission.

## OPTOELECTRONIC PROPERTIES AND DEVICES

Having isolated and fabricated individual, few-layer nanoribbon FETs, their optical properties have been studied by illuminating the transport channel with a laser of a particular wavelength. The photoresponse of individual TiS$_3$ nanoribbon FETs was reported recently by Island *et al.*.[53] Figure 9(a) shows a model representation of a TiS$_3$ device studied in the report, illuminated with a red laser. The study consisted of photocurrent measurements in a vacuum probe station equipped with 5 lasers ($\lambda$ = 532 nm, 640 nm, 808 nm, 885 nm, and 940 nm). A beam spot size of 200 μm diameter was used for all wavelengths. This ensured that the entire transport channel was illuminated reducing contributions from the bolometric and photothermoelectric effects due to sharp thermal gradients. Figure 9(b) shows a transfer curve for a typical TiS$_3$ nanoribbon FET with a ribbon thickness of 22 nm, at a bias voltage of 500 mV in dark conditions (black line) and upon illumination from a laser ($P$ = 500 μW, $\lambda$ = 532 nm, solid green line). The inset shows the drain current as a function of bias voltage at a gate voltage of -40 V. In both cases there is a clear increase in drain current upon illumination. The photocurrent ($I_{ph}$) is calculated as the difference in the drain current in dark conditions and under illumination and at a gate voltage of +40V, the device produces a photocurrent of 100 nA. In the OFF state at a gate voltage of -40 V, the device produces a photocurrent of ~40 nA. In Figure 9(c) the current-voltage (I-V) characteristics at a back gate voltage of -40 V for increasing powers of the incident laser ($\lambda$ = 640 nm) are shown. It was found that the drain current increased sublinearly with laser power. This is characteristic of photodetectors in which photocurrent generation is sustained by trapping and recombination of photogenerated carriers. Due to the large surface to volume ratio in nanostructured photodetectors, photocurrent generation is susceptible to charge trapping from surface or interface traps.[160-162] The trapping of photogenerated carriers leads to a sublinear response of the photocurrent as a function of increased laser power. This can lead to increased gains as trapped carriers can further "gate" the flakes but this gain comes at the expense of slower photoresponse.[163] The gain of the photodetector is estimated by calculating the responsivity ($R = I_{ph}/P$, where $P$ is the power of the laser which has been scaled by the ratio of the area of the device (195 nm × 470 nm) to the area of the laser spot (diameter 200 μm). At an incident power of 100 nW, the responsivity reaches 1030 A/W (2910 A/W for best device measured, see ref. [53]). This responsivity is several orders of magnitude larger than intrinsic graphene and even $MoS_2$ photodetectors which achieve a responsivity up to 880 A/W.[164, 165]

By varying the wavelength of the laser excitation, the spectral response was reported in the same study. Figure 9(e) shows the measured photocurrent ($V_b$ = 1V, $V_g$ = -40 V) as a function of laser wavelength ($P$ = 500 μW). An estimate of the bandgap energy of the photodetector can be made by making a linear fit of the measured photocurrent as a function of wavelength (dashed line in Figure 9(e)). A bandgap energy of roughly 1.2 eV is estimated which agrees reasonably well with the bulk-like value of ~1 eV.

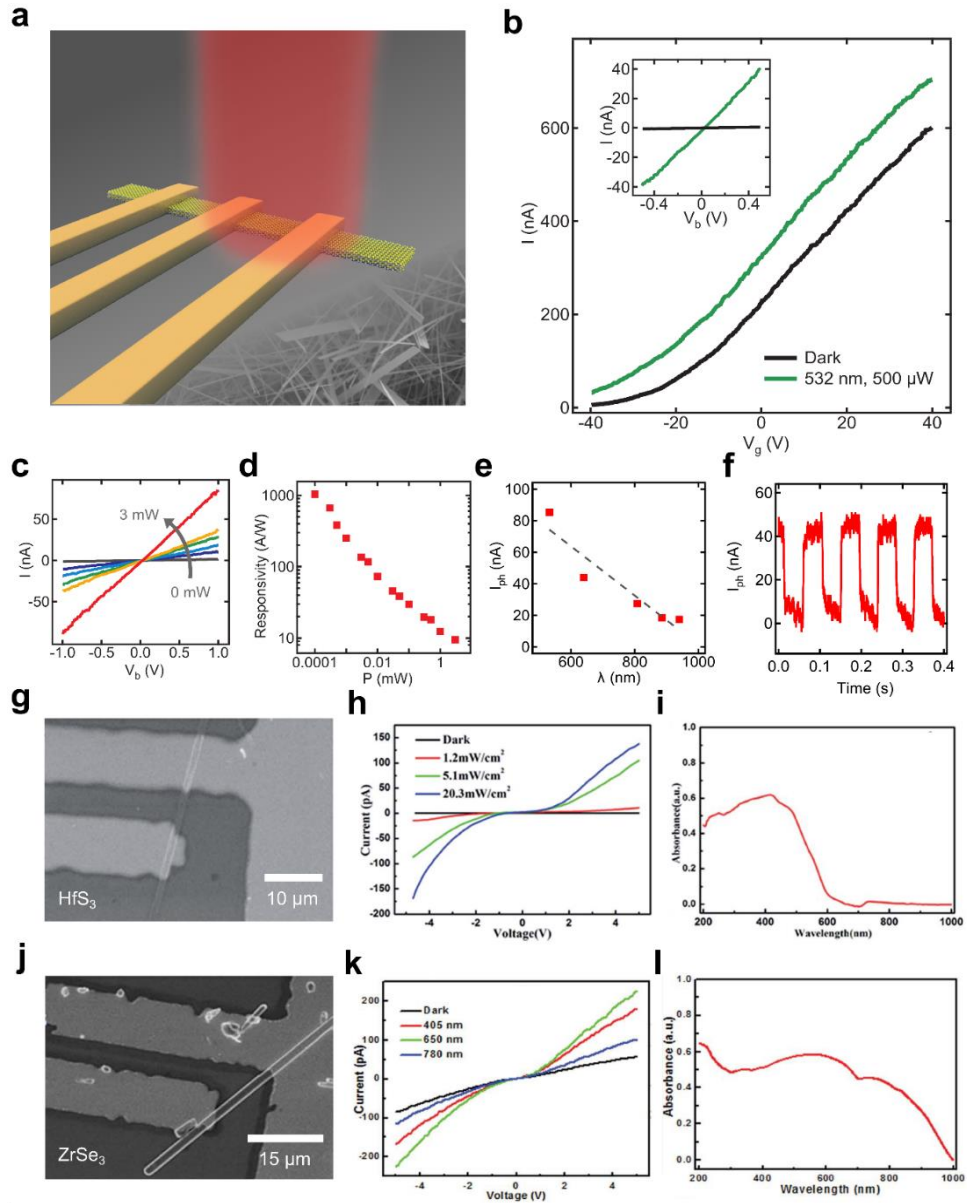

Figure 9. (a) Model representation of a TiS$_3$ nanoribbon photodetector illuminated with a red laser. (b) Transfer curve ($V_b$ = 500 mV) for a TiS$_3$ nanoribbon photodetector in dark conductions (black curve) and upon illumination (green curve). Inset shows the current voltage characteristics at $V_g$ = -40 V for the same laser excitation. (c) Current-voltage characteristics at $V_g$ = −40 V in dark (black solid line) and under 640 nm excitation for increasing laser powers up to 3 mW. (d) Log-log plot of the responsivity (R) as a function of excitation power ($\lambda$ = 640 nm). (e) Photocurrent as a function of illumination wavelength. (f) Photocurrent response using a 10 Hz mechanically modulated optical excitation ($\lambda$ = 640 nm, P = 500 μW). Panels (a-f) have been adapted from ref. [53] with permission. (g) A typical SEM image of an individual HfS$_3$ nanobelt photodetector. (h) The I–V characteristics of an HfS$_3$ nanobelt photodetector illuminated with different laser powers. (i) The UV-VIS absorption spectrum of HfS$_3$ nanobelts. Panels (g-i) are adapted from Ref. [75] with permission. (j) A typical SEM image of a single ZrSe$_3$ nanobelt photodetector. (k) The I–V characteristics of the single ZrSe$_3$ nanobelt photodetector under illumination with different wavelengths and dark conditions. (l) The UV-VIS absorption spectrum of ZrSe$_3$ nanobelts. Panels (j-l) are adapted from Ref. [166] with permission.

The response time of the TiS$_3$ photodetector was characterized by modulating the intensity of the excitation laser using a mechanical chopper. Figure 9(f) shows the measured photocurrent as a function of time at a 10 Hz modulation of the laser. Repeated modulation of the excitation laser shows that the photocurrent is stable over several cycles. The rise and fall time of the photodetector were estimated to be a few to several milliseconds. By changing the chopper frequency, the cutoff frequency was estimated to be 1000 Hz. The fast response and high cutoff frequency of TiS$_3$ nanoribbons make them an ideal material for nanostructured photodetectors.

The remaining group IV trichalcogenides have larger bandgaps ranging from ~1.5 eV to ~3 eV and could be alternatives to GaAs (1.3 eV) and wide bandgap UV photodetectors.[167] While progress in photodetection with these materials is not as extensive as with TiS$_3$, the materials have been characterized and simple photodetectors have been fabricated. Different photodetector characteristics for trichalcogenide-based materials are listed in table 4.

ZrS$_3$ is a layered p-type semiconductor from the TMTCs family with an important anisotropic character. Zr atoms lie at the center of distorted trigonal prisms of S atoms, arranged in successive packing in infinite chains along the crystal b-axis; chains are held together in the c-axis by van der Waals forces.[87, 168] ZrS$_3$, in a nanobelt morphology, has been used in the fabrication of photodetectors showing photosensitivity to wavelengths from 405 nm to $\lambda$ = 780 nm. The devices present ON/OFF current ratios of about 1.5 with dark current of 109 pA and response times around 390/800 ms (rise time/decay time).[76] The responsivity measured for this material is 500 mA W$^{-1}$ with a cutoff wavelength for strong photoresponsivity around 850 nm.[75, 76]

HfS$_3$ is another p-type semiconductor from the TMTCs family with the same crystal structure as ZrS$_3$ but where Hf atoms replace Zr.[168] It presents a bulk direct bandgap of 3.1 eV.[169] As has been accomplished with ZrS$_3$, HfS$_3$ nanobelt photodetectors have been fabricated and studied with 405 nm wavelength light illumination.[75] Figure 9(g) shows an SEM image of a typical device. The measured ON/OFF current ratio is 337.5 with a dark current of 0.04 pA, responsivity of 110 mA W$^{-1}$ and rise/decay time lower than 400 ms. Wavelength-dependent current-voltage characteristics of HfS$_3$ nanobelt photodetectors are shown in Figure 9(h-i).

ZrSe$_3$ is a n-type semiconductor which presents an indirect bandgap at 1.54 eV in bulk, while HfSe$_3$ (see Figure 9(j)) presents a direct bandgap at 2.31 eV.[128, 170, 171] Both materials have been used for visible light photodetection with individual nanobelt devices under illumination with wavelengths of 405 nm, 650 nm and 780 nm, obtaining ON/OFF ratios of 2 and 2.5 for ZrSe$_3$ and HfSe$_3$, respectively.[166] Current-voltage characteristics of the devices fabricated with ZrSe$_3$ in dark conditions and under illumination with different wavelengths are shown in Figure 9(k-i), as well as the absorption spectra as a function of the light wavelength.

A deeper understanding of the optical properties of layered low-dimensional materials usually requires the investigation of optical transitions and excitonic bound states in their band structure. Working in this line, Molina-Mendoza *et al.* have studied the electronic bandgap of TiS$_3$ by means of scanning tunnelling spectroscopy (STS) on TiS$_3$ at room temperature and determined the exciton binding energy by comparing the electronic bandgap with the optical bandgap.[47] STS measurements were carried out on mechanically exfoliated TiS$_3$ transferred onto a Au(111) substrate. Tunneling current-bias voltages curves are used to determine the valence and conduction band values with respect to the Fermi level (Figure 10(a)). As the STS measurements are performed in ambient conditions, a statistical analysis is used to overcome the issue of perturbations in the STS spectra due to diffusion and thermal drift. This statistical analysis consists of building a 2D histogram with a set of IVs (205 traces in Figure 10(a)) and then extracting a line profile along zero current (dashed line in Figure 10(a)) to represent the number of counts as a function of the bias voltage (Figure 10(b)). The valence and conduction bands are then determined as 50% of the highest count along zero current (red lines in Figure 10(b)).[172] The obtained values for the valence and conduction bands are $E_{VB} = -0.64 \pm 0.06$ eV and $E_{CB} = 0.47 \pm 0.06$ eV, respectively, yielding an electronic bandgap value of $E_{g,el}^{exp} = E_{CB} - E_{VB} + E_t = 1.2 \pm 0.08$ eV (where $E_t$ is the thermal contribution). This value is then compared with the optical bandgap measured by photoelectrochemical measurements, where they obtain an optical bandgap of $E_{g,op}^{exp} = 1.07 \pm 0.01$ eV (Figure 10(c)). Finally, the exciton binding energy is determined as the difference in energy between the electronic bandgap and the optical bandgap: $E_{exc}^{exp} = E_{g,el}^{exp} - E_{g,op}^{exp} = 130$ meV. These results are in good agreement with random phase approximation (RPA) and Bethe-Salpeter equation (BSE) calculations, where they find an electronic bandgap of $E_{g,el}^{th} = 1.15$ eV, an optical bandgap of $E_{g,op}^{th} = 1.05$ eV and exciton binding energy of $E_{exc}^{th} = 100$ meV.

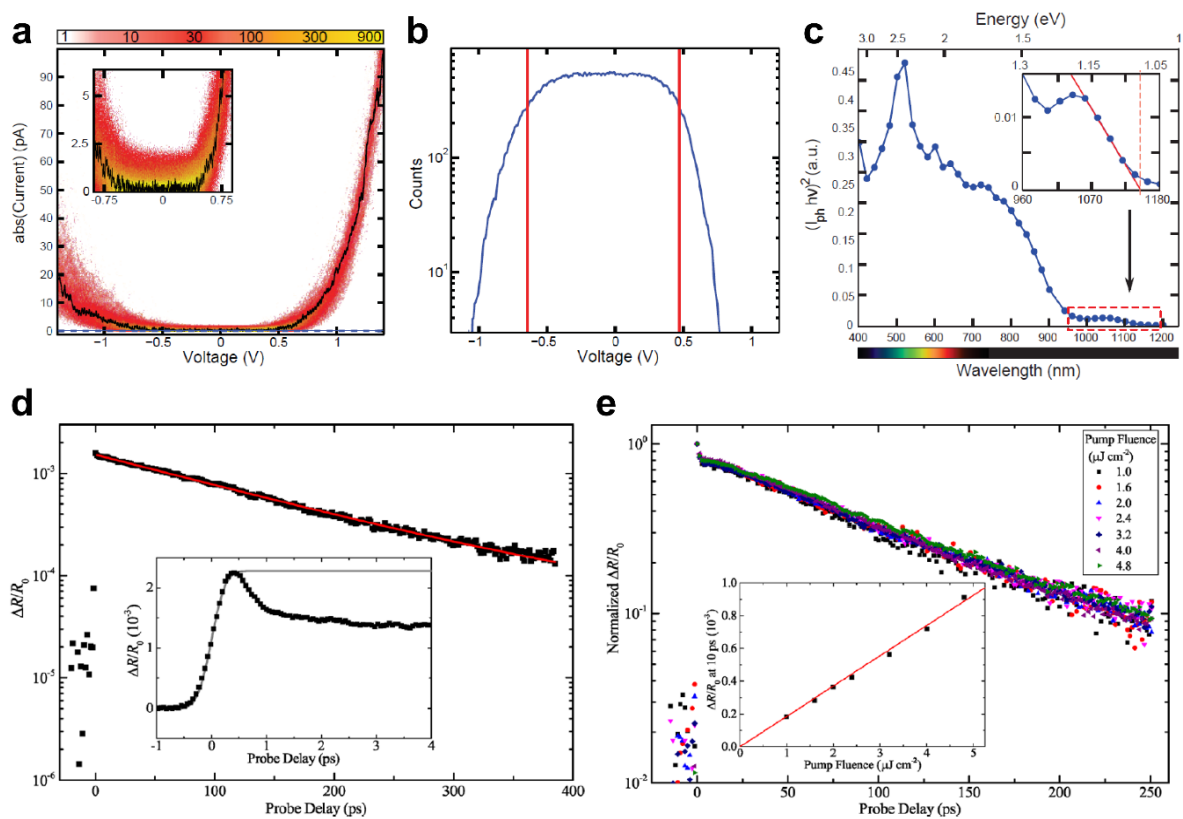

Figure 10. (a) Colormap histogram including 205 STS current–voltage characteristics (current in absolute value). A representative current–voltage curve is plotted in solid black line on top. Inset: Zoom around zero current in the colormap histogram. (b) 1D histogram extracted from a profile along the dashed line (zero current) in the 2D histogram in panel (a). The vertical red lines indicate the valence and conduction band values ($-0.64 \pm 0.06$ eV and $0.47 \pm 0.06$ eV, respectively). (c) Photocurrent density as a function of light wavelength (energy). The optical bandgap energy ($1.07 \pm 0.01$ eV) is determined by the point where the linear fit cuts the zero photocurrent density (highlighted in the inset). Panels a-c) adapted from ref. [47] with permission. (d) Differential reflection signal of a 950 nm probe pulse as a function of the probe delay after the sample is excited by a 395 nm pump pulse. The red curve is a single-exponential fit with a time constant of 140 ps. The inset shows the signal at early delays. The gray curve is the integral of a Gaussian function with a width of 0.47 ps. (e) Normalized differential reflection signal of a 950 nm probe pulse as a function of the probe delay after the sample is excited by a 395 nm pump pulse. The values of the pump fluence are indicated in the label. The inset shows the magnitude of the signal, reflected by the signal at a probe delay of 10 ps, as a function of the pump fluence. Panels (d-e) reprinted (adapted) with permission from ref. [173]. Copyright 2016 American Chemical Society.

More recently, Cui *et al.* have reported time resolved measurements on individual TiS$_3$ nanoribbons probing the intrinsic dynamic response of excited photocarriers.[173] Photocarriers were excited with a 295 nm pump pulse with a time duration of 200 fs. The dynamics were then monitored using a probe pulse of 950 nm with a 300 fs duration (Figure 10(d-e)). They found that thermalization and energy relaxation of the photocarriers occurs on a time scale faster than 0.5 ps. After this transient, the measurements give a short timescale for exciton formation and an overall exciton lifetime of 140 ps. Using spatially and temporally resolved measurements, the authors also estimate an exciton diffusion coefficient of 1.2 cm$^2$/s, a diffusion length of 130 nm, and an exciton mobility of 50 cm$^2$/Vs.

Table 4. MX$_3$ based photodetectors characteristics

| Material | Measurements conditions | | | | Responsivity | Rise time | | Ref |
|---|---|---|---|---|---|---|---|---|
| | $V_{sd}$ (V) | $V_g$ (V) | λ (nm) | P (nW cm$^{-2}$) | R (mA W$^{-1}$) | τ (ms) | Spectral range | |
| TiS$_3$ | 1 | 40 | 640 | 3·10$^{-1}$ | 2.9·10$^6$ | 4 | λ < 1100nm | 53 |
| ZrS$_3$ | 5 | - | 405 | 5·10$^{11}$ | 500 | 13·10$^3$ | λ < 850 nm | 76 |
| ZrSe$_3$ | 5 | - | 650 | 3.8·10$^6$ | 12 | 400 | λ < 650 nm | 166 |
| HfS$_3$ | 5 | - | 405 | 1.2 | 110 | 400 | λ < 650 nm | 75 |
| HfSe$_3$ | 5 | - | 532 | 5.3·10$^6$ | 530 | 400 | λ < 650 nm | 166 |

# ANISOTROPY AND LINEAR DICHROISM

In this section we review the recent progress on results related to the most interesting property of the MX$_3$ class of materials. The reduced symmetry in the crystal structure of the TMTCs sets them apart from graphene and the TMDCs and their largely isotropic in-plane electrical and optical properties. This property initially attracted researchers in the early 1970s and the various materials in the family have enjoyed growing interest since then. Besides studies of 1D conduction and CDW mentioned in the introduction, several groups have investigated anisotropic properties of TMTCs in bulk form. Notable contributions include electrical transport measurements on TiS$_3$[125, 137, 138], thermodynamic[129, 174] and optical[126, 175] measurements of ZrTe$_3$, and polarized reflectance comparisons between NbSe$_3$ and TaSe$_3$.[176]

Recently, the anisotropic electrical and optical properties of few-layer TiS$_3$ flakes have been investigated.[27] Figure 11(a) shows an atomic force microscopy (AFM) image of a representative device from the study. Electrodes are patterned parallel to one edge of the flake and at increments of 30 degrees in order to measure electrical conduction along the b-axis (high mobility axis), a-axis (low mobility axis), and increments in between. Figure 11(b) shows a polar plot of the room temperature conductance for different pairs of electrodes (two terminal measurements) for different back gate voltages (-40V, 0V, and 40V). Along the b-axis, a mobility of 25 cm$^2$/Vs is measured (extracted from the linear regime transconductance) and an anisotropic ratio of $G_{max}/G_{min}$ = 2.1 ($\mu_b/\mu_a$ = 2.3) is extracted. These results are in agreement with theoretical calculations that predict a lower effective mass (higher mobility) for electrons along the b-axis than for electrons along the a-axis.[25] Furthermore, the anisotropy in the conduction increases with decreased temperature. Figure 11(c) shows the same measurement at 25 K. The anisotropy ratios increase to $G_{max}/G_{min}$ = 4.4 ($\mu_b/\mu_a$ = 7.6). In addition to this large electrical anisotropy, the optical properties of TiS$_3$ nanoribbons show marked anisotropy as well (linear dichroism). Figure 11(d) shows the transmittance of the red, blue, and green channels of polarized optical measurements of a single TiS$_3$ nanoribbon. The polarization of the optical filter is indicated next to the optical images shown on the outside of the polar plot. The transmittance is seen to vary greatly with the direction of the polarization, indeed, the transmittance is nearly quenched for polarizations along the b-axis. These findings are supported by calculations of the absorption and transmittance of TiS$_3$ as a function of energy (Figure 11(e)) and electric field direction (inset of Figure 11(e)). Absorption along the b-axis (dotted line) is much greater than along the a-axis (solid line). The transmittance (inset of Figure 11(e)) follows the same qualitative response as found in the measurements. The red channel (1.9 eV) has the largest transmittance and the blue channel (2.72 eV) has the smallest. The transmittance is also seen to vary greatly with excitation direction.

These results confirm earlier theoretical predictions based on first-principles calculations which reported the anisotropic optical absorption over a broader range of energies.[25] Figure 11(f) shows the calculated optical absorption coefficients for TiS$_3$ for polarizations along the XX direction (a-axis, across the chains), YY direction (b-axis, along the chains) and the ZZ direction (c-axis, perpendicular to the covalently bonded layers). The optical absorption at low energies is greatest for polarizations along the b-axis, in agreement with calculations in Fig. 11(e). Additionally, the study reported optical absorption for TiSe$_3$, shown in Fig. 11(g), which also presents an interesting anisotropic absorption with an inversion of the absorption coefficients around 2 eV for the XX and YY directions.

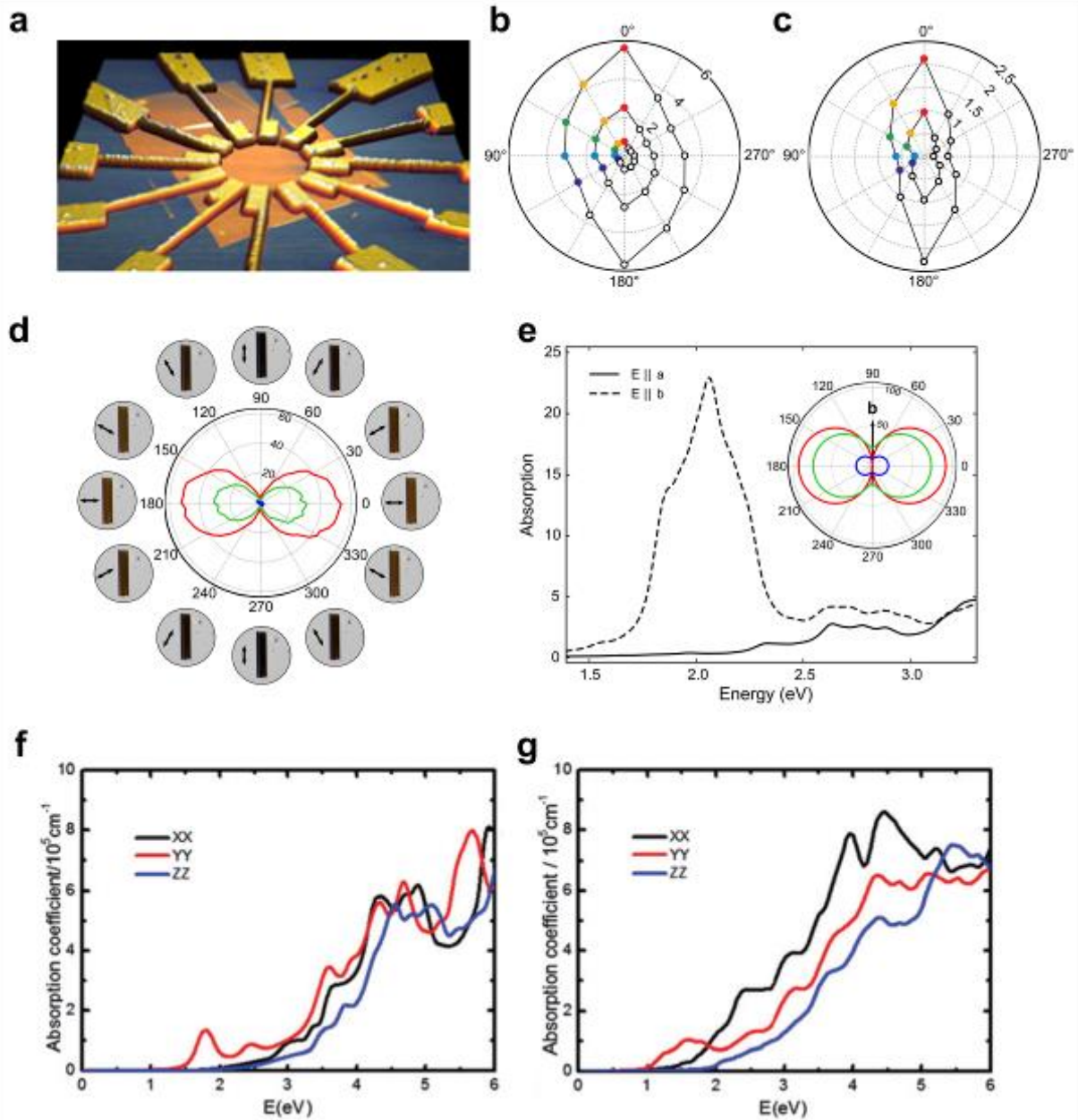

Figure 11. (a) AFM image of a few-layer TiS$_3$ flake with Ti/Au electrodes. (b) Polar plot of the room temperature conductance (μS) measured between all 12 electrodes at a backgate voltage of -40 V, 0 V, and 40 V. (c) Polar plot at 25 K of the conductance (μS) measured between all 12 electrodes at a backgate voltage of -40 V, 0 V, and 40 V. (d) Transmittance of the red, green, and blue channels as a function of the excitation polarization angle. (e) Calculated absorption spectra when the field is aligned parallel to the b-axis (dashed line) and a-axis (solid line). The inset shows the transmittance in the a–b plane for energies red (1.9 eV), green (2.4 eV), and blue (2.72 eV) excitations. This figure adapted from Ref. [27] with permission. (f) Calculated optical absorption coefficient of TiS$_3$ for polarizations along the a-axis (XX), b-axis (YY), and c-axis (ZZ). (g) Calculated optical absorption coefficient

foe TiSe$_3$ for polarizations along the a-axis (XX), b-axis (YY), and c-axis (ZZ). Panels (f-g) are adapted from ref. [25] with permission.

Pant et al. have demonstrated the first anisotropic light emission from a TMTC using photoluminescence measurements on ZrS$_3$.[28] Figure 12(a) shows the PL emission as a function of azimuthal flake angle with respect to the excitation laser. The PL is seen to vary greatly with angle and energy. By integrating the PL peak area for different azimuthal angles, figure 12(b) clearly shows that the PL response is maximum when the excitation is aligned parallel to the crystal chains (b-axis). This is in contrast with the more commonly studied 2D semiconductors such as MoS$_2$ (red dots in Figure 12(b)) which do not present a modulation of PL intensity with excitation direction. Figure 12(c) shows that the dichroic emission is preserved in thin exfoliated specimens as well. The authors find that the polarization anisotropy is ultimately weaker than strictly 1D nanowire/nanotube materials which is attributed to finite interactions between the 1D chains in the TMTC family materials.

Shortly after this, the characteristics of the lattice vibrations in TiS$_3$ as a function of pressure were reported by Wu et al.[177] Figure 12(d-e) show the Raman vibrational peaks of TiS$_3$ as a function of azimuthal angle in the normal (excitation and detection aligned parallel to the crystal chains, b-axis) and orthogonal (excitation perpendicular to the detection polarization which is parallel to the crystal chains) configurations. The measured angle here is measured between the b-axis and the detection direction. The intensity of the peaks is seen to vary as a function of azimuthal angle. For increasing pressure, most of the Raman modes are shown to stiffen (Figure 12(f)) as expected but, interestingly, the Ag$^{S-S}$ mode shows an unconventional negative pressure dependence. This is attributed to a softening of the S-S vibration through an increased bond distance upon increased pressure.

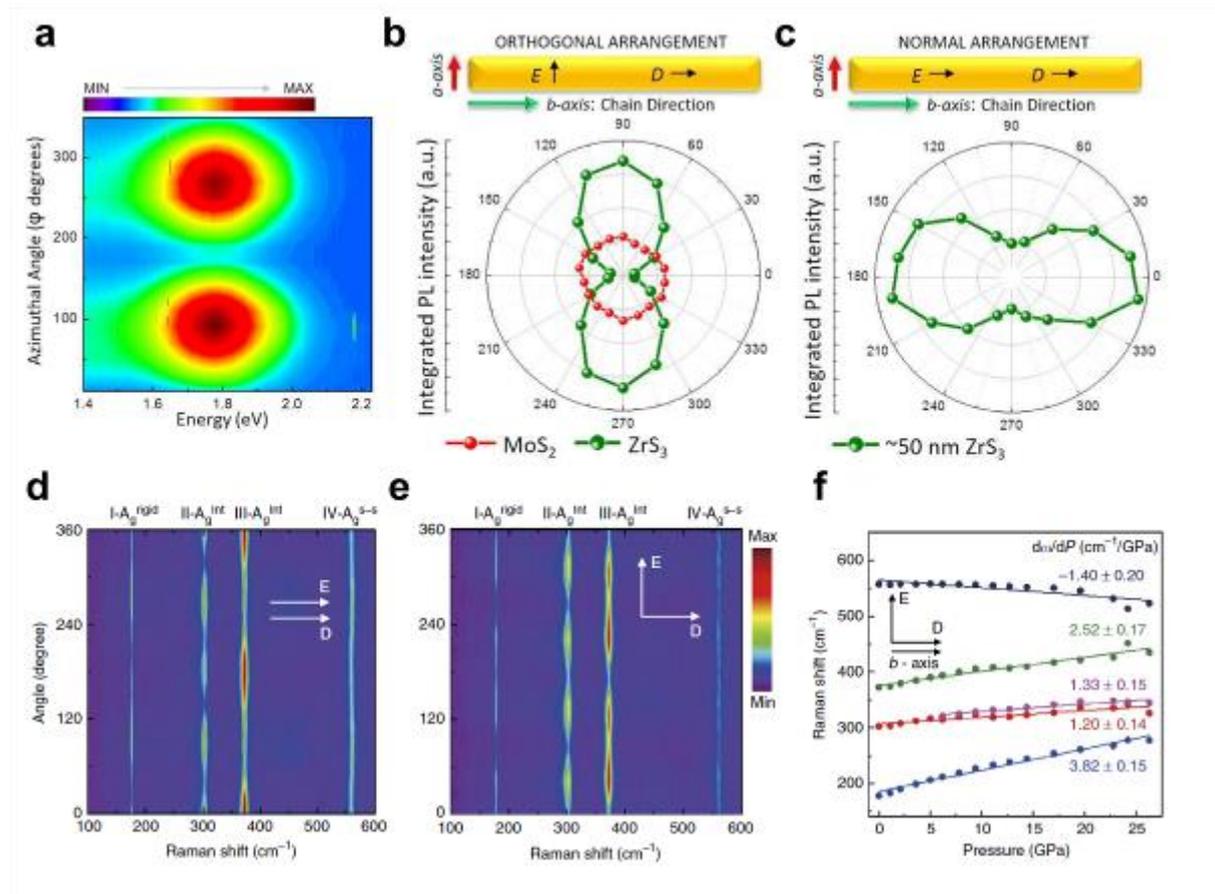

Figure 12: (a) Contour plot of the angle dependence of the PL emission from a ~50 nm thick ZrS$_3$ flake in orthogonal arrangement. (b) Polar plot of integrated peak area with azimuthal flake angle (φ) in orthogonal arrangement (c) Polar plot of the integrated peak area with azimuthal angle in the normal arrangement. Panels (a-c) adapted from Ref. [28] with permission. (d) Angle-resolved Raman spectroscopy (2D contour plots) for all the Raman peaks in the (d) normal (E parallel D) and (e) orthogonal (E perpendicular D) configuration for TiS$_3$. E and D represent the excitation direction and detection direction, respectively. (f) Evolution of Raman spectrum in orthogonal configuration displaying similar pressure-dependent trends and dω/dP values. Panels (d-f) adapted from Ref. [177] with permission.

## SUMMARY AND OUTLOOK

We have reviewed the application of the TMTCs as novel materials for next generation electronic and optoelectronic devices. We discussed the various growth methods that have been utilized over the years to produce copious amounts of MX$_3$ material in the form of a powder of nanoribbons. We summarized characterizations of the bulk material by optical absorption and X-ray diffraction. The electronic structure of the TMTCs was reviewed including recent developments in theoretical calculations within DFT and many-body methods. Recent research involving mechanical and chemical exfoliation of individual nanoribbons was discussed. We then reviewed the progress in fabrication and measurement of field-effect transistors (FET) and electrical devices employing TMTCs. The TMTCs are found to make good electrical interconnects and high current density transistors. The optical properties and devices incorporating TMTCs was reviewed. TMTCs are found to be remarkable materials for high gain photodetection.

Research on atomically thin TMTCs has only just begun. There are many directions for improvement and innovation in the growth, theory, and device optimization of MX$_3$s. In the next few paragraphs we summarize the most pressing issues regarding this class of materials and suggest several directions for further studies.

From a growth perspective, certainly the most interesting prospective direction is the synthesis of large area, single layer MX3s which has greatly advanced research on other 2D materials such as graphene[178-180], MoS$_2$[181-183], and other TMDCs[184, 185]. In this respect, control over the morphology of TiS$_3$ has facilitated exfoliation of the material down to a single layer[46] but large area synthesis would preclude the need for additional fabrication steps. Control over dopants (vancancies, interstitial, and substitutional dopants) would facilitate growth of p-type materials and materials that achieve ambipolar transport in FET devices.

TMTCs currently present important challenges for theory and are forming an interesting playground for ab-initio methods. One of the first challenges to address is the nature of the gap for the bulk materials: invariably, theory, in the form of standard DFT, more refined hybrid functionals, and non-self consistent *GW* calculations, predicts an indirect gap, while there is clear evidence that, at least for TiS$_3$, the gap is direct at the $\Gamma$ point. Understanding the nature of the gap will reinforce subsequent theoretical models. Coinciding with challenges in growth of MX$_3$s, a second major theoretical task is the clear identification of suitable dopants to control the electronic gap. Here, theory can help identify the ideal dopants to guide synthesis and experimental studies. Moreover, among the TMTCs, surely TiS$_3$ is the material attracting the most attention and the remaining materials have yet to be investigated in detail. Here again, theory can facilitate experiments in performing a pre-screening for the best materials, offering experimentalists a database of properties to select from.

In terms of electronic transport devices, an outstanding challenge is fabricating cleaner and higher quality FETs with mobilities that are closer to predicted values. TiS$_3$ for instance has a predicted b-axis mobility of 10000 cm$^2$/Vs (ref. [26]) but the highest mobility achieved to date is 73 cm$^2$/Vs (ref. [46]). This experimental mobility was achieved at the cost of a lower ON/OFF ratio due to sulphur vancancies. Improved mobilities were reported using conventional dielectrics (up to 43 cm$^2$/Vs) which better preserved ON/OFF ratios (up to 7000) but still fell well below the predicted values.[52] Improvements in synthesis and defect control will contribute to better starting materials for fabricating high quality devices[186]. Certainly these materials will also benefit from being combined with other 2D materials to create ultraclean heterostructures using boron nitride as a substrate and graphite as a local gate. This has proven extremely useful in fabricating high mobility graphene[187, 188], black phosphorus[189-191], and MoS$_2$[192, 193] devices. With improvements in device quality, theories of reduced backscattering in anisotropic devices can be tested.[32-34] An interesting prospect in high quality single layer MX$_3$ devices is the study of an anisotropic quantum hall[194] and anisotropic fractional quantum hall effect[195-198].

The optical properties of the TMTCs offer many directions for future studies. Simple photodetectors display ultrahigh responsivisties (up to 3000 A/W) without optimization[53]. Increased photoconductive gain could be achieved through the realization of hybrid devices that benefit from coupling to quantum dots[199] or functionalized molecules[200]. While time resolved measurements have been performed on bulk-like nanoribbons (3 μm thick)[173], atomically thin crystals have yet to be investigated and it remains to be seen whether the TMTCs present similar many body interactions (exciton-exciton annihilation) as the TMDCs in atomically thin samples[201-205]. Perhaps the most intriguing direction though is the application of $MX_3$s in creating on-chip polarizers[36-39], polarization sensitive photodetectors[40-42], and devices with polarized light emission[28, 43, 44].

Finally, the TMTCs must be contrasted with the recently (re)discovered anisotropic 2D materials black phosphorus and rhenium disulphide ($ReS_2$). Initial studies of atomically thin TMTCs, such as $TiS_3$, show that single and few layers are stable in air and do not suffer from ambient degradation which has hindered initial studies on black phosphorus[206-210]. $ReS_2$, on the other hand, is also stable in air and has presented exceptional anisotropic properties[211-213]. $ReS_2$ however is a single unique member of the TMDCs with a bandgap of 1.4 eV and does not offer the wide variety of metals, semiconductors (bandgaps ranging from 0.2 eV to 2 eV), and superconductors with similar anisotropic properties which are waiting to be explored.

**Acknowledgements** This work was supported by the Netherlands Organization for Scientific Research (NWO/FOM). A.J.M-M. acknowledges the financial support of Ministerio de Ciencia e Innovación (MICINN) (Spain) through the scholarship BES-2012-057346. R. D'A. and R. B. acknowledge financial support by the DYN-XC-TRANS (Grant No. FIS2013-43130-P), NanoTHERM (Grant No. CSD2010-00044), and SElecT-DFT (FIS2016-79464-P) of the Ministerio de Economia y Competitividad (MINECO), and Grupo Consolidado UPV/EHU del Gobierno Basco (Grant No. IT578-13). R. B. acknowledges the financial support of the Ministerio de Educacion, Cultura y Deporte (Grant No. FPU12/01576). AC-G acknowledges financial support from the European Commission under the Graphene Flagship, contract CNECTICT-604391, from the MINECO (Ramón y Cajal 2014 program, RYC-2014-01406) and from the MICINN (MAT2014-58399-JIN). MIRE Group thanks MINECO (MAT2015-65203R) for financial support. E. Flores also acknowledges the Mexican National Council for Science and Technology (CONACyT).

1. Wang, Q. H.; Kalantar-Zadeh, K.; Kis, A.; Coleman, J. N.; Strano, M. S. *Nature Nanotechnology* **2012,** 7, (11), 699-712.
2. Butler, S. Z.; Hollen, S. M.; Cao, L.; Cui, Y.; Gupta, J. A.; Gutierrez, H. R.; Heinz, T. F.; Hong, S. S.; Huang, J.; Ismach, A. F. *ACS nano* **2013,** 7, (4), 2898-2926.
3. Chhowalla, M.; Shin, H. S.; Eda, G.; Li, L.-J.; Loh, K. P.; Zhang, H. *Nature chemistry* **2013,** 5, (4), 263-275.
4. Buscema, M.; Island, J. O.; Groenendijk, D. J.; Blanter, S. I.; Steele, G. A.; van der Zant, H. S.; Castellanos-Gomez, A. *Chemical Society Reviews* **2015,** 44, (11), 3691-3718.
5. Castellanos-Gomez, A. *The journal of physical chemistry letters* **2015,** 6, (21), 4280-4291.
6. Ling, X.; Wang, H.; Huang, S.; Xia, F.; Dresselhaus, M. S. *Proceedings of the National Academy of Sciences* **2015,** 112, (15), 4523-4530.
7. Ferrari, A. C.; Bonaccorso, F.; Fal'Ko, V.; Novoselov, K. S.; Roche, S.; Bøggild, P.; Borini, S.; Koppens, F. H.; Palermo, V.; Pugno, N. *Nanoscale* **2015,** 7, (11), 4598-4810.
8. Yagmurcukardes, M.; Peeters, F.; Senger, R.; Sahin, H. *Applied Physics Reviews* **2016,** 3, (4), 041302.
9. Congxin, X.; Jingbo, L. *Journal of Semiconductors* **2016,** 37, (5), 051001.
10. Lv, R.; Robinson, J. A.; Schaak, R. E.; Sun, D.; Sun, Y.; Mallouk, T. E.; Terrones, M. *Accounts of chemical research* **2014,** 48, (1), 56-64.
11. Bhimanapati, G. R.; Lin, Z.; Meunier, V.; Jung, Y.; Cha, J.; Das, S.; Xiao, D.; Son, Y.; Strano, M. S.; Cooper, V. R. *ACS nano* **2015,** 9, 11509-11539.

12. Xu, M.; Liang, T.; Shi, M.; Chen, H. *Chemical reviews* **2013,** 113, (5), 3766-3798.
13. Castellanos-Gomez, A. *Nature Photonics* **2016,** 10, (4), 202-204.
14. Novoselov, K.; Geim, A.; Morozov, S.; Jiang, D.; Grigorieva, M. I. K. I. V.; Dubonos, S.; Firsov, A. *Nature* **2005,** 438, (7065), 197-200.
15. Britnell, L.; Ribeiro, R.; Eckmann, A.; Jalil, R.; Belle, B.; Mishchenko, A.; Kim, Y.-J.; Gorbachev, R.; Georgiou, T.; Morozov, S. *Science* **2013,** 340, (6138), 1311-1314.
16. Mak, K. F.; Lee, C.; Hone, J.; Shan, J.; Heinz, T. F. *Physical Review Letters* **2010,** 105, (13), 136805.
17. Pu, J.; Yomogida, Y.; Liu, K.-K.; Li, L.-J.; Iwasa, Y.; Takenobu, T. *Nano Letters* **2012,** 12, (8), 4013-4017.
18. Lee, G.-H.; Yu, Y.-J.; Cui, X.; Petrone, N.; Lee, C.-H.; Choi, M. S.; Lee, D.-Y.; Lee, C.; Yoo, W. J.; Watanabe, K. *ACS Nano* **2013,** 7, (9), 7931-7936.
19. Geim, A.; Grigorieva, I. *Nature* **2013,** 499, (7459), 419-425.
20. Yu, W. J.; Li, Z.; Zhou, H.; Chen, Y.; Wang, Y.; Huang, Y.; Duan, X. *Nature materials* **2013,** 12, (3), 246-252.
21. Yu, W. J.; Liu, Y.; Zhou, H.; Yin, A.; Li, Z.; Huang, Y.; Duan, X. *Nature Nanotechnology* **2013,** 8, (12), 952-958.
22. Klein, A.; Tiefenbacher, S.; Eyert, V.; Pettenkofer, C.; Jaegermann, W. *Physical Review B* **2001,** 64, (20), 205416.
23. Jin, W.; Yeh, P.-C.; Zaki, N.; Zhang, D.; Sadowski, J. T.; Al-Mahboob, A.; van Der Zande, A. M.; Chenet, D. A.; Dadap, J. I.; Herman, I. P. *Physical Review Letters* **2013,** 111, (10), 106801.
24. Lebègue, S.; Björkman, T.; Klintenberg, M.; Nieminen, R.; Eriksson, O. *Physical Review X* **2013,** 3, (3), 031002.
25. Jin, Y.; Li, X.; Yang, J. *Physical Chemistry Chemical Physics* **2015,** 17, 18665-18669.
26. Dai, J.; Zeng, X. C. *Angewandte Chemie* **2015,** 54, (26), 7572-7576.
27. Island, J. O.; Biele, R.; Barawi, M.; Clamagirand, J. M.; Ares, J. R.; Sánchez, C.; van der Zant, H. S.; Ferrer, I. J.; D'Agosta, R.; Castellanos-Gomez, A. *Scientific reports* **2016,** 6, 22214.
28. Pant, A.; Torun, E.; Chen, B.; Bhat, S. S.; Fan, X.; Wu, K.; Wright, D. P.; Peeters, F.; Soignard, E.; Sahin, H. *Nanoscale* **2016,** 8, 16259-16265.
29. Dai, J.; Li, M.; Zeng, X. C. *Wiley Interdisciplinary Reviews: Computational Molecular Science* **2016,** 6, 211-222.
30. Kong, W.; Bacaksiz, C.; Chen, B.; Wu, K.; Blei, M.; Fan, X.; Shen, Y.; Sahin, H.; Wright, D.; Narang, D. S.; Tongay, S. *Nanoscale* **2017,** 9, 4175-4182.
31. Silva-Guillén, J.; Canadell, E.; Ordejón, P.; Guinea, F.; Roldán, R. *arXiv preprint arXiv:1704.05824* **2017**.
32. Bufler, F.; Fichtner, W. *IEEE Transactions on Electron Devices* **2003,** 50, (2), 278-284.
33. Abudukelimu, A.; Kakushima, K.; Ahmet, P.; Geni, M.; Tsutsui, K.; Nishiyama, A.; Sugii, N.; Natori, K.; Hattori, T.; Iwai, H. In *The effect of isotropic and anisotropic scattering in drain region of ballistic channel diode*, Solid-State and Integrated Circuit Technology (ICSICT), 2010 10th IEEE International Conference on, 2010; IEEE: pp 1247-1249.
34. Bufler, F.; Keith, S.; Meinerzhagen, B., Anisotropic Ballistic In—Plane Transport of Electrons in Strained Si. In *Simulation of Semiconductor Processes and Devices 1998*, Springer: 1998; pp 239-242.
35. Liu, E.; Fu, Y.; Wang, Y.; Feng, Y.; Liu, H.; Wan, X.; Zhou, W.; Wang, B.; Shao, L.; Ho, C.-H. *Nature communications* **2015,** 6, 6991.
36. Sasagawa, K.; Shishido, S.; Ando, K.; Matsuoka, H.; Noda, T.; Tokuda, T.; Kakiuchi, K.; Ohta, J. *Optics express* **2013,** 21, (9), 11132-11140.
37. Shishido, S.; Noda, T.; Sasagawa, K.; Tokuda, T.; Ohta, J. *Japanese Journal of Applied Physics* **2011,** 50, (4S), 04DL01.
38. Liao, Y.-L.; Zhao, Y. *Optical and Quantum Electronics* **2014,** 46, (5), 641-647.
39. Guillaumée, M.; Dunbar, L.; Santschi, C.; Grenet, E.; Eckert, R.; Martin, O.; Stanley, R. *Applied Physics Letters* **2009,** 94, (19), 193503.
40. Yuan, H.; Liu, X.; Afshinmanesh, F.; Li, W.; Xu, G.; Sun, J.; Lian, B.; Curto, A. G.; Ye, G.; Hikita, Y. *Nature nanotechnology* **2015,** 10, (8), 707-713.

41. Zhang, E.; Wang, P.; Li, Z.; Wang, H.; Song, C.; Huang, C.; Chen, Z.-G.; Yang, L.; Zhang, K.; Lu, S. *ACS nano* **2016,** 10, (8), 8067-8077.
42. Nanot, S.; Cummings, A. W.; Pint, C. L.; Ikeuchi, A.; Akiho, T.; Sueoka, K.; Hauge, R. H.; Léonard, F.; Kono, J. *Scientific reports* **2013,** 3, 1335.
43. Matioli, E.; Brinkley, S.; Kelchner, K. M.; Hu, Y.-L.; Nakamura, S.; DenBaars, S.; Speck, J.; Weisbuch, C. *Light: Science & Applications* **2012,** 1, (8), e22.
44. Singer, S.; Mecklenburg, M.; White, E.; Regan, B. *Physical Review B* **2011,** 83, (23), 233404.
45. Ferrer, I. J.; Ares, J. R.; Clamagirand, J. M.; Barawi, M.; Sánchez, C. *Thin Solid Films* **2012,** 535, 398-401.
46. Island, J. O.; Barawi, M.; Biele, R.; Almazán, A.; Clamagirand, J. M.; Ares, J. R.; Sánchez, C.; van der Zant, H. S.; Álvarez, J. V.; D'Agosta, R.; Ferrer, I. J.; Castellanos-Gomez, A. *Advanced Materials* **2015,** 27, (16), 2595-2601.
47. Molina-Mendoza, A. J.; Barawi, M.; Biele, R.; Flores, E.; Ares, J. R.; Sánchez, C.; Rubio-Bollinger, G.; Agraït, N.; D'Agosta, R.; Ferrer, I. J.; Castellanos-Gomez, A. *Advanced Electronic Materials* **2015,** 1, (9), 1500126.
48. Kang, J.; Wang, L.-W. *Physical Chemistry Chemical Physics* **2016,** 18, (22), 14805-14809.
49. Roldán, R.; Castellanos-Gomez, A.; Cappelluti, E.; Guinea, F. *Journal of Physics: Condensed Matter* **2015,** 27, (31), 313201.
50. Island, J. O.; Kuc, A.; Diependaal, E. H.; Bratschitsch, R.; van der Zant, H. S.; Heine, T.; Castellanos-Gomez, A. *Nanoscale* **2016,** 8, (5), 2589-2593.
51. Conley, H. J.; Wang, B.; Ziegler, J. I.; Haglund Jr, R. F.; Pantelides, S. T.; Bolotin, K. I. *Nano Letters* **2013,** 13, (8), 3626-3630.
52. Lipatov, A.; Wilson, P. M.; Shekhirev, M.; Teeter, J. D.; Netusil, R.; Sinitskii, A. *Nanoscale* **2015,** 7, (29), 12291-12296.
53. Island, J. O.; Buscema, M.; Barawi, M.; Clamagirand, J. M.; Ares, J. R.; Sánchez, C.; Ferrer, I. J.; Steele, G. A.; van der Zant, H. S.; Castellanos-Gomez, A. *Advanced Optical Materials* **2014,** 2, (7), 641-645.
54. Solyom, J. *Advances in Physics* **1979,** 28, (2), 201-303.
55. Horovitz, B.; Gutfreund, H.; Weger, M. *Physical Review B* **1975,** 12, (8), 3174.
56. Monceau, P., *Electronic Properties of Inorganic Quasi-One-Dimensional Compounds: Part II*. D. Reidel: Dordrecht, 1985; Vol. 2.
57. Peierls, R. *Annalen der Physik* **1930,** 396, (2), 121-148.
58. Monceau, P. *Advances in Physics* **2012,** 61, (4), 325-581.
59. Frohlich, H. In *On the theory of superconductivity: the one-dimensional case*, Proceedings of the Royal Society of London A: Mathematical, Physical and Engineering Sciences, 1954; The Royal Society: 1954; pp 296-305.
60. Ong, N.; Monceau, P. *Physical Review B* **1977,** 16, (8), 3443.
61. Monceau, P.; Peyrard, J.; Richard, J.; Molinie, P. *Physical Review Letters* **1977,** 39, (3), 161.
62. Chianelli, R. R.; Dines, M. B. *Inorganic Chemistry* **1975,** 14, (10), 2417-2421.
63. Murphy, D.; Trumbore, F. *Journal of Crystal Growth* **1977,** 39, (1), 185-199.
64. Ōnuki, Y.; Inada, R.; Tanuma, S.; Yamanaka, S.; Kamimura, H. *Solid State Ionics* **1983,** 11, (3), 195-201.
65. Wu, J.; Wang, D.; Liu, H.; Lau, W.-M.; Liu, L.-M. *RSC Advances* **2015,** 5, (28), 21455-21463.
66. Matsuyama, T.; Hayashi, A.; Hart, C. J.; Nazar, L. F.; Tatsumisago, M. *Journal of The Electrochemical Society* **2016,** 163, (8), A1730-A1735.
67. Chen, Z.; Cummins, D.; Reinecke, B. N.; Clark, E.; Sunkara, M. K.; Jaramillo, T. F. *Nano letters* **2011,** 11, (10), 4168-4175.
68. Denholme, S.; Dobson, P.; Weaver, J.; MacLaren, I.; Gregory, D. *International Journal of Nanotechnology* **2012,** 9, (1-2), 23-38.
69. Gorochov, O.; Katty, A.; Le Nagard, N.; Levy-Clement, C.; Schleich, D. *Materials Research Bulletin* **1983,** 18, (1), 111-118.
70. Ferrer, I.; Maciá, M.; Carcelén, V.; Ares, J.; Sánchez, C. *Energy Procedia* **2012,** 22, 48-52.
71. Xie, J.; Wang, R.; Bao, J.; Zhang, X.; Zhang, H.; Li, S.; Xie, Y. *Inorganic Chemistry Frontiers* **2014,** 1, (10), 751-756.

72. Barawi, M.; Flores, E.; Ferrer, I.; Ares, J.; Sánchez, C. *Journal of Materials Chemistry A* **2015,** 3, (15), 7959-7965.
73. Flores, E.; Ares, J.; Ferrer, I.; Sánchez, C. *physica status solidi (RRL)-Rapid Research Letters* **2016,** 10, (11), 802-806.
74. Tao, Y.-R.; Chen, J.-Q.; Wu, J.-J.; Wu, Y.; Wu, X.-C. *Journal of Alloys and Compounds* **2016,** 658, 6-11.
75. Xiong, W.-W.; Chen, J.-Q.; Wu, X.-C.; Zhu, J.-J. *J. Mater. Chem. C* **2014,** 2, (35), 7392-7395.
76. Tao, Y.-R.; Wu, X.-C.; Xiong, W.-W. *Small* **2014,** 10, (23), 4905-4911.
77. Guilmeau, E.; Berthebaud, D.; Misse, P. R.; Hébert, S.; Lebedev, O. I.; Chateigner, D.; Martin, C.; Maignan, A. *Chemistry of Materials* **2014,** 26, (19), 5585-5591.
78. Misse, P.; Berthebaud, D.; Lebedev, O. I.; Maignan, A.; Guilmeau, E. *Materials* **2015,** 8, (5), 2514-2522.
79. Liang, K.; Jacobson, A.; Chianelli, R.; Betts, F. *Journal of Non-Crystalline Solids* **1980,** 35, 1249-1254.
80. Murugesan, T.; Gopalakrishnan, J. In *Amorphous MoS3 and A x MoS3 (A= Li or Na; 0< x< 4)*, Proceedings of the Indian Academy of Sciences-Chemical Sciences, 1982; Springer: 1982; pp 7-13.
81. Chianelli, R. *International Reviews in Physical Chemistry* **1982,** 2, (2), 127-165.
82. Cramer, S.; Liang, K.; Jacobson, A.; Chang, C.; Chianelli, R. *Inorganic Chemistry* **1984,** 23, (9), 1215-1221.
83. Liang, K.; Cramer, S.; Johnston, D.; Chang, C.; Jacobson, A.; Chianelli, R. *Journal of Non-Crystalline Solids* **1980,** 42, (1-3), 345-356.
84. Afanasiev, P. *Comptes Rendus Chimie* **2008,** 11, (1), 159-182.
85. Morales-Guio, C. G.; Hu, X. *Accounts of chemical research* **2014,** 47, (8), 2671-2681.
86. Merki, D.; Fierro, S.; Vrubel, H.; Hu, X. *Chemical Science* **2011,** 2, (7), 1262-1267.
87. Furuseth, S.; Brattas, L.; Kjekshus, A. *Acta Chemica Scandinavica* **1975,** 29, 623-631.
88. Bullett, D. *Journal of Physics C: Solid State Physics* **1979,** 12, (2), 277-281.
89. Gopalakrishnan, J.; Nanjundaswamy, K. *Bulletin of Materials Science* **1983,** 5, (3-4), 287-306.
90. Wilson, J. *Physical Review B* **1979,** 19, (12), 6456-6468.
91. Clamagirand, J. M. Gas/solid reactions in the formation of metal sulfide thin films for thermoelectric applications. PhD Thesis, Autonomous University of Madrid, Spain, 2016.
92. Brattas, L.; Kjekshus, A. *Acta Chemica Scandinavica* **1972,** 26, (9), 3441-3449.
93. Grimmeiss, H.; Rabenau, A.; Hahn, H.; Ness, P. *Zeitschrift für Elektrochemie, Berichte der Bunsengesellschaft für physikalische Chemie* **1961,** 65, (9), 776-783.
94. Takahashi, S.; Sambongi, T.; Okada, S. *Le Journal de Physique Colloques* **1983,** 44, (C3), C3-1733-C3-1736.
95. Itkis, M.; Nad, F. Y.; Levy, F. *Synthetic Metals* **1991,** 43, (3), 3969-3972.
96. Wang, Z. Z.; Monceau, P.; Salva, H.; Roucau, C.; Guemas, L.; Meerschaut, A. *Physical Review B* **1989,** 40, (17), 11589-11593.
97. Chaussy, J.; Haen, P.; Lasjaunias, J.; Monceau, P.; Waysand, G.; Waintal, A.; Meerschaut, A.; Molinié, P.; Rouxel, J. *Solid State Communications* **1976,** 20, (8), 759-763.
98. Roucau, C.; Ayroles, R.; Monceau, P.; Guemas, L.; Meerschaut, A.; Rouxel, J. *physica status solidi (a)* **1980,** 62, (2), 483-493.
99. Nichols, J.; Weerasooriya, C. S.; Brill, J. *Journal of Physics: Condensed Matter* **2010,** 22, (33), 334224.
100. Thompson, A.; Zettl, A.; Grüner, G. *Physical Review Letters* **1981,** 47, (1), 64.
101. Sambongi, T.; Yamamoto, M.; Tsutsumi, K.; Shiozaki, Y.; Yamaya, K.; Abe, Y. *Journal of the Physical Society of Japan* **1977,** 42, (4), 1421-1422.
102. McTaggart, F. K.; Wadsley, A. *Australian Journal of Chemistry* **1958,** 11, (4), 445-457.
103. Haraldsen, H. K., Arne; Røst, Erling; Steffensen, Arne. *Acta Chem. Scand* **1963,** 17, (5), 1283-1292.
104. Blitz, W.; Ehrlich, P. *Zeitschrift für anorganische und allgemeine Chemie* **1937,** 234, (2), 97-116.
105. Murphy, D. W.; Trumbore, F. A. *Journal of The Electrochemical Society* **1976,** 123, (7), 960-964.
106. Endo, K.; Ihara, H.; Watanabe, K.; Gonda, S.-I. *Journal of Solid State Chemistry* **1982,** 44, (2), 268-272.
107. Jin, H.; Cheng, D.; Li, J.; Cao, X.; Li, B.; Wang, X.; Liu, X.; Zhao, X. *Solid State Sciences* **2011,** 13, (5), 1166-1171.
108. Khumalo, F.; Hughes, H. *Physical Review B* **1980,** 22, (4), 2078-2088.

109. Srivastava, S. K.; Avasthi, B. N. *J Mater Sci* **1992,** 27, (14), 3693-3705.
110. Levy, F.; Berger, H. *Journal of Crystal Growth* **1983,** 61, (1), 61-68.
111. Bensalem, A.; Schleich, D. *Materials research bulletin* **1988,** 23, (6), 857-868.
112. Chang, H.; Schleich, D. *Journal of Solid State Chemistry* **1992,** 100, (1), 62-70.
113. Huang, L.; Tang, K.; Yang, Q.; Shen, G.; Jia, S. *Materials research bulletin* **2004,** 39, (7), 1083-1089.
114. Kikkawa, S.; Ogawa, N.; Koizumi, M.; Onuki, Y. *Journal of Solid State Chemistry* **1982,** 41, (3), 315-322.
115. Kikkawa, S.; Shinya, K.; Koizumi, M. *Journal of Solid State Chemistry* **1982,** 41, (3), 323-328.
116. Wu, X. C.; Tao, Y. R.; Gao, Q. X. *Nano Research* **2009,** 2, (7), 558-564.
117. Jeannin, Y. *Ann. Chim.* **1962,** 7, 57-83.
118. Kikkawa, S.; Koizumi, M.; Yamanaka, S.; Onuki, Y.; Tanuma, S. *physica status solidi (a)* **1980,** 61, (1), K55-K57.
119. Endo, K.; Ihara, H.; Watanabe, K.; Gonda, S.-I. *Journal of Solid State Chemistry* **1981,** 39, (2), 215-218.
120. Hsieh, P.-L.; Jackson, C.; Grüner, G. *Solid state communications* **1983,** 46, (7), 505-507.
121. Finkman, E.; Fisher, B. *Solid state communications* **1984,** 50, (1), 25-28.
122. Fleet, M. E.; Harmer, S. L.; Liu, X.; Nesbitt, H. W. *Surface science* **2005,** 584, (2), 133-145.
123. Ma, J.; Liu, X.; Cao, X.; Feng, S.; Fleet, M. E. *European journal of inorganic chemistry* **2006,** 2006, (3), 519-522.
124. Gorlova, I. G.; Pokrovskii, V. Y. *JETP letters* **2009,** 90, (4), 295-298.
125. Gorlova, I.; Pokrovskii, V. Y.; Zybtsev, S.; Titov, A.; Timofeev, V. *Journal of Experimental and Theoretical Physics* **2010,** 111, (2), 298-303.
126. Bayliss, S.; Liang, W. *Journal of Physics C: Solid State Physics* **1981,** 14, (26), L803-L807.
127. Redon, A. *Electrochimica acta* **1985,** 30, (10), 1365-1369.
128. Kurita, S.; Staehli, J. L.; Guzzi, M.; Lévy, F. *Physica B+C* **1981,** 105, (1–3), 169-173.
129. Felser, C.; Finckh, E.; Kleinke, H.; Rocker, F.; Tremel, W. *Journal of Materials Chemistry* **1998,** 8, (8), 1787-1798.
130. Gard, P.; Cruege, F.; Sourisseau, C.; Gorochov, O. *Journal of Raman spectroscopy* **1986,** 17, (3), 283-288.
131. Gard, P.; Sourisseau, C.; Gorochov, O. *physica status solidi (b)* **1987,** 144, (2), 885-901.
132. Ye, C. *Science of Advanced Materials* **2010,** 2, (3), 365-377.
133. Umar, A.; Kim, S.; Lee, Y.-S.; Nahm, K.; Hahn, Y. *Journal of Crystal Growth* **2005,** 282, (1), 131-136.
134. Burton, W.-K.; Cabrera, N.; Frank, F. *Philosophical Transactions of the Royal Society of London A: Mathematical, Physical and Engineering Sciences* **1951,** 243, (866), 299-358.
135. Kubelka, P.; Munk, F. *Z. Tech. Phys* **1931,** 12, 593-601.
136. Cullity, B. D., *Elements of X-ray Diffraction*. Addison Wesley: New York, 1978.
137. Gorlova, I.; Zybtsev, S.; Pokrovskii, V. Y.; Bolotina, N.; Verin, I.; Titov, A. *Physica B: Condensed Matter* **2012,** 407, (11), 1707-1710.
138. Gorlova, I. G.; Zybtsev, S. G. e.; Pokrovskii, V. Y. *JETP Letters* **2014,** 100, (4), 256-261.
139. Gorlova, I.; Zybtsev, S.; Pokrovskii, V. Y.; Bolotina, N.; Gavrilkin, S. Y.; Tsvetkov, A. Y. *Physica B: Condensed Matter* **2015,** 460, 11-15.
140. Osada, K.; Bae, S.; Tanaka, M.; Raebiger, H.; Shudo, K.; Suzuki, T. *The Journal of Physical Chemistry C* **2016,** 120, (8), 4653-4659.
141. Stolyarov, M. A.; Liu, G.; Bloodgood, M. A.; Aytan, E.; Jiang, C.; Samnakay, R.; Salguero, T. T.; Nika, D. L.; Bozhilov, K. N.; Balandin, A. A. *Nanoscale* **2016,** 8, 15774-15782.
142. Poltarak, P.; Artemkina, S.; Bulavchenko, A.; Podlipskaya, T. Y.; Fedorov, V. *Russian Chemical Bulletin* **2015,** 64, (8), 1850-1856.
143. Fedorov, V. E.; Artemkina, S. B.; Grayfer, E. D.; Naumov, N. G.; Mironov, Y. V.; Bulavchenko, A. I.; Zaikovskii, V. I.; Antonova, I. V.; Komonov, A. I.; Medvedev, M. V. *Journal of Materials Chemistry C* **2014,** 2, (28), 5479-5486.
144. Artemkina, S. B.; Podlipskaya, T. Y.; Bulavchenko, A. I.; Komonov, A. I.; Mironov, Y. V.; Fedorov, V. E. *Colloids and Surfaces A: Physicochemical and Engineering Aspects* **2014,** 461, 30-39.
145. Abdulsalam, M.; Joubert, D. P. *The European Physical Journal B* **2015,** 88, (7), 1-11.

146. Abdulsalam, M.; Joubert, D. P. *physica status solidi (b)* **2016,** 253, (5), 868-874.
147. Li, M.; Dai, J.; Zeng, X. C. *Nanoscale* **2015,** 7, (37), 15385-15391.
148. Bullett, D. *Journal of Solid State Chemistry* **1980,** 33, (1), 13-16.
149. Perluzzo, G.; Lakhani, A.; Jandl, S. *Solid State Communications* **1980,** 35, (3), 301-304.
150. Zwick, A.; Renucci, M.; Kjekshus, A. *Journal of Physics C: Solid State Physics* **1980,** 13, (30), 5603-5614.
151. Patel, K.; Prajapati, J.; Vaidya, R.; Patel, S. *Bulletin of Materials Science* **2005,** 28, (5), 405-410.
152. Patel, K.; Prajapati, J.; Vaidya, R.; Patel, S. In *Study of optical and electrical properties for ZrS3 single crystals*, INDIAN JOURNAL OF PHYSICS AND PROCEEDINGS OF THE INDIAN ASSOCIATION FOR THE CULTIVATION OF SCIENCE, 2005; INDIAN ASSOC CULTIVATION SCIENCE INDIAN J PHYSICS, JADAVPUR, KOLKATA 700 032, INDIA: pp 373-376.
153. Iyikanat, F.; Sahin, H.; Senger, R. T.; Peeters, F. *The Journal of Physical Chemistry C* **2015,** 119, (19), 10709-10715.
154. Kang, J.; Sahin, H.; Ozaydin, H. D.; Senger, R. T.; Peeters, F. M. *Physical Review B* **2015,** 92, (7), 075413.
155. Biele, R.; Flores, E.; Ares, J. R.; Sanchez, C.; Ferrer, I. J.; Rubio-Bollinger, G.; Castellanos-Gomez, A.; D'Agosta, R. *arXiv preprint arXiv:1509.00532* **2015**.
156. Aierken, Y.; Çakır, D.; Peeters, F. M. *Physical Chemistry Chemical Physics* **2016,** 18, (21), 14434-14441.
157. Kang, J.; Sahin, H.; Peeters, F. M. *Physical Chemistry Chemical Physics* **2015,** 17, (41), 27742-27749.
158. Liu, G.; Rumyantsev, S.; Bloodgood, M. A.; Salguero, T. T.; Shur, M.; Balandin, A. A. *Nano Letters* **2016,** 17, 377-383.
159. Molina-Mendoza, A. J.; Island, J. O.; Paz, W. S.; Clamagirand, J. M.; Ares, J. R.; Flores, E.; Leardini, F.; Sánchez, C.; Agraït, N.; Rubio-Bollinger, G.; Van der Zant, H. S.; Ferrer, I. J.; Palacios, J. J.; Castellanos-Gomez, A. *Advanced Functional Materials* **2017,** 27, 1605647.
160. Soci, C.; Zhang, A.; Xiang, B.; Dayeh, S. A.; Aplin, D.; Park, J.; Bao, X.; Lo, Y.-H.; Wang, D. *Nano Letters* **2007,** 7, (4), 1003-1009.
161. González-Posada, F.; Songmuang, R.; Den Hertog, M.; Monroy, E. *Nano Letters* **2011,** 12, (1), 172-176.
162. Furchi, M. M.; Polyushkin, D. K.; Pospischil, A.; Mueller, T. *Nano Letters* **2014,** 14, (11), 6165-6170.
163. Island, J. O.; Blanter, S. I.; Buscema, M.; van der Zant, H. S.; Castellanos-Gomez, A. *Nano letters* **2015,** 15, (12), 7853-7858.
164. Xia, F.; Mueller, T.; Golizadeh-Mojarad, R.; Freitag, M.; Lin, Y.-m.; Tsang, J.; Perebeinos, V.; Avouris, P. *Nano Letters* **2009,** 9, (3), 1039-1044.
165. Lopez-Sanchez, O.; Lembke, D.; Kayci, M.; Radenovic, A.; Kis, A. *Nature Nanotechnology* **2013,** 8, (7), 497-501.
166. Xiong, W.-W.; Chen, J.-Q.; Wu, X.-C.; Zhu, J.-J. *Journal of Materials Chemistry C* **2015,** 3, (9), 1929-1934.
167. Monroy, E.; Omnès, F.; Calle, F. *Semiconductor Science and Technology* **2003,** 18, (4), R33-R51.
168. Canadell, E.; Thieffry, C.; Mathey, Y.; Whangbo, M. H. *Inorganic Chemistry* **1989,** 28, (15), 3043-3047.
169. Schairer, W.; Shafer, M. *physica status solidi (a)* **1973,** 17, (1), 181-184.
170. Hankare, P. P.; Asabe, M. R.; Kokate, A. V.; Delekar, S. D.; Sathe, D. J.; Mulla, I. S.; Chougule, B. K. *Journal of Crystal Growth* **2006,** 294, (2), 254-259.
171. Cingolani, A.; Lugarà, M.; Lévy, F. *Physica Scripta* **1988,** 37, (3), 389-391.
172. Castellanos-Gomez, A.; Wojtaszek, M.; Arramel; Tombros, N.; van Wees, B. J. *Small* **2012,** 8, (10), 1607-1613.
173. Cui, Q.; Lipatov, A.; Wilt, J. S.; Bellus, M. Z.; Zeng, X. C.; Wu, J.; Sinitskii, A.; Zhao, H. *ACS Applied Materials & Interfaces* **2016,** 8, (28), 18334-18338.
174. O'Neal, T.; Brill, J.; Ganguly, B.; Xiang, X.-D.; Minton, G. *Synthetic Metals* **1989,** 31, (2), 215-223.
175. Herr, S.; Brill, J. *Synthetic metals* **1986,** 16, (3), 283-290.
176. Geserich, H.; Scheiber, G.; Lévy, F.; Monceau, P. *Physica B+ C* **1986,** 143, (1-3), 174-176.
177. Wu, K.; Torun, E.; Sahin, H.; Chen, B.; Fan, X.; Pant, A.; Wright, D. P.; Aoki, T.; Peeters, F. M.; Soignard, E. *Nature Communications* **2016,** 7, 12952.
178. Kim, K. S.; Zhao, Y.; Jang, H.; Lee, S. Y.; Kim, J. M.; Kim, K. S.; Ahn, J.-H.; Kim, P.; Choi, J.-Y.; Hong, B. H. *nature* **2009,** 457, (7230), 706-710.

179. Li, X.; Cai, W.; An, J.; Kim, S.; Nah, J.; Yang, D.; Piner, R.; Velamakanni, A.; Jung, I.; Tutuc, E. *Science* **2009,** 324, (5932), 1312-1314.
180. Bae, S.; Kim, H.; Lee, Y.; Xu, X.; Park, J.-S.; Zheng, Y.; Balakrishnan, J.; Lei, T.; Kim, H. R.; Song, Y. I. *Nature nanotechnology* **2010,** 5, (8), 574-578.
181. Lee, Y. H.; Zhang, X. Q.; Zhang, W.; Chang, M. T.; Lin, C. T.; Chang, K. D.; Yu, Y. C.; Wang, J. T. W.; Chang, C. S.; Li, L. J. *Advanced Materials* **2012,** 24, (17), 2320-2325.
182. Liu, K.-K.; Zhang, W.; Lee, Y.-H.; Lin, Y.-C.; Chang, M.-T.; Su, C.-Y.; Chang, C.-S.; Li, H.; Shi, Y.; Zhang, H. *Nano letters* **2012,** 12, (3), 1538-1544.
183. Zhan, Y.; Liu, Z.; Najmaei, S.; Ajayan, P. M.; Lou, J. *Small* **2012,** 8, (7), 966-971.
184. Gao, Y.; Liu, Z.; Sun, D.-M.; Huang, L.; Ma, L.-P.; Yin, L.-C.; Ma, T.; Zhang, Z.; Ma, X.-L.; Peng, L.-M. *Nature communications* **2015,** 6, 8569.
185. Shi, Y.; Zhang, H.; Chang, W.-H.; Shin, H. S.; Li, L.-J. *MRS Bulletin* **2015,** 40, (07), 566-576.
186. Lin, Z.; Carvalho, B. R.; Kahn, E.; Lv, R.; Rao, R.; Terrones, H.; Pimenta, M. A.; Terrones, M. *2D Materials* **2016,** 3, (2), 022002.
187. Dean, C. R.; Young, A. F.; Meric, I.; Lee, C.; Wang, L.; Sorgenfrei, S.; Watanabe, K.; Taniguchi, T.; Kim, P.; Shepard, K. *Nature nanotechnology* **2010,** 5, (10), 722-726.
188. Gannett, W.; Regan, W.; Watanabe, K.; Taniguchi, T.; Crommie, M.; Zettl, A. *Applied Physics Letters* **2011,** 98, (24), 242105.
189. Li, L.; Yang, F.; Ye, G. J.; Zhang, Z.; Zhu, Z.; Lou, W.; Zhou, X.; Li, L.; Watanabe, K.; Taniguchi, T. *Nature nanotechnology* **2016,** 11, 593-597.
190. Li, L.; Ye, G. J.; Tran, V.; Fei, R.; Chen, G.; Wang, H.; Wang, J.; Watanabe, K.; Taniguchi, T.; Yang, L. *Nature nanotechnology* **2015,** 10, (7), 608-613.
191. Chen, X.; Wu, Y.; Wu, Z.; Han, Y.; Xu, S.; Wang, L.; Ye, W.; Han, T.; He, Y.; Cai, Y. *Nature communications* **2015,** 6, 7315.
192. Wu, Z.; Xu, S.; Lu, H.; Khamoshi, A.; Liu, G.-B.; Han, T.; Wu, Y.; Lin, J.; Long, G.; He, Y.; Cai, Y.; Yao, Y.; Zhang, F.; Wang, N. *Nature Communications* **2016,** 7, 12955.
193. Lee, G.-H.; Cui, X.; Kim, Y. D.; Arefe, G.; Zhang, X.; Lee, C.-H.; Ye, F.; Watanabe, K.; Taniguchi, T.; Kim, P. *ACS nano* **2015,** 9, (7), 7019-7026.
194. Einevoll, G.; Lütken, C. *Physical Review B* **1993,** 48, (15), 11492.
195. Ghazaryan, A.; Chakraborty, T. *Physical Review B* **2015,** 92, (16), 165409.
196. Johri, S.; Papić, Z.; Schmitteckert, P.; Bhatt, R.; Haldane, F. *New Journal of Physics* **2016,** 18, (2), 025011.
197. Balram, A. C.; Jain, J. *Physical Review B* **2016,** 93, (7), 075121.
198. Mueed, M.; Kamburov, D.; Hasdemir, S.; Pfeiffer, L.; West, K.; Baldwin, K.; Shayegan, M. *Physical Review B* **2016,** 93, (19), 195436.
199. Konstantatos, G.; Badioli, M.; Gaudreau, L.; Osmond, J.; Bernechea, M.; de Arquer, F. P. G.; Gatti, F.; Koppens, F. H. L. *Nat Nano* **2012,** 7, (6), 363-368.
200. Schmidt, H.; Giustiniano, F.; Eda, G. *Chemical Society Reviews* **2015,** 44, (21), 7715-7736.
201. Sun, D.; Rao, Y.; Reider, G. A.; Chen, G.; You, Y.; Brézin, L.; Harutyunyan, A. R.; Heinz, T. F. *Nano letters* **2014,** 14, (10), 5625-5629.
202. Kar, S.; Su, Y.; Nair, R.; Sood, A. *ACS nano* **2015,** 9, (12), 12004-12010.
203. Kumar, N.; Cui, Q.; Ceballos, F.; He, D.; Wang, Y.; Zhao, H. *Physical Review B* **2014,** 89, (12), 125427.
204. Yuan, L.; Huang, L. *Nanoscale* **2015,** 7, (16), 7402-7408.
205. Mouri, S.; Miyauchi, Y.; Toh, M.; Zhao, W.; Eda, G.; Matsuda, K. *Physical Review B* **2014,** 90, (15), 155449.
206. Castellanos-Gomez, A.; Vicarelli, L.; Prada, E.; Island, J. O.; Narasimha-Acharya, K.; Blanter, S. I.; Groenendijk, D. J.; Buscema, M.; Steele, G. A.; Alvarez, J. *2D Materials* **2014,** 1, (2), 025001.
207. Wood, J. D.; Wells, S. A.; Jariwala, D.; Chen, K.-S.; Cho, E.; Sangwan, V. K.; Liu, X.; Lauhon, L. J.; Marks, T. J.; Hersam, M. C. *Nano letters* **2014,** 14, (12), 6964-6970.
208. Ziletti, A.; Carvalho, A.; Campbell, D. K.; Coker, D. F.; Neto, A. C. *Physical review letters* **2015,** 114, (4), 046801.
209. Island, J. O.; Steele, G. A.; van der Zant, H. S.; Castellanos-Gomez, A. *2D Materials* **2015,** 2, (1), 011002.

210. Kim, J.-S.; Liu, Y.; Zhu, W.; Kim, S.; Wu, D.; Tao, L.; Dodabalapur, A.; Lai, K.; Akinwande, D. *Scientific Reports* **2015,** 5, 8989.
211. Lin, Y.-C.; Komsa, H.-P.; Yeh, C.-H.; Björkman, T.; Liang, Z.-Y.; Ho, C.-H.; Huang, Y.-S.; Chiu, P.-W.; Krasheninnikov, A. V.; Suenaga, K. *ACS Nano* **2015,** 9, (11), 11249-11257.
212. Chenet, D. A.; Aslan, O. B.; Huang, P. Y.; Fan, C.; van der Zande, A. M.; Heinz, T. F.; Hone, J. C. *Nano Letters* **2015,** 15, (9), 5667-5672.
213. Jariwala, B.; Voiry, D.; Jindal, A.; Chalke, B. A.; Bapat, R.; Thamizhavel, A.; Chhowalla, M.; Deshmukh, M.; Bhattacharya, A. *Chemistry of Materials* **2016,** 28, (10), 3352-3359.